\documentclass[onecolumn]{elsart3p}

\newcommand{\Jnature}{Nature (London)}
\newcommand{\Jnatphys}{Nat. Phys.}
\newcommand{\Jnatcomm}{Nat. Comm.}

\newcommand{\Jscience}{Science}
\newcommand{\Jsciadv}{Science Advances}

\newcommand{\Jprx}{Phys. Rev. X}
\newcommand{\Jprl}{Phys. Rev. Lett.}
\newcommand{\Jpr}{Phys. Rev.}
\newcommand{\Jpra}{Phys. Rev. A}
\newcommand{\Jprb}{Phys. Rev. B}

\newcommand{\Jrmp}{Rev. Mod. Phys.}

\newcommand{\Jepl}{Europhys. Lett.}
\newcommand{\Jnjp}{New J. Phys.}

\newcommand{\JphysA}{J. Phys. A}

\newcommand{\Jprocroysoc}{Proc. Roy. Soc. A: Math. Phys. Eng. Sci.}

\newcommand{\Jijthphys}{Int. J. Theor. Phys.}

\newcommand{\Jphysrep}{Phys. Rep.}
\newcommand{\JRepProgPhys}{Rep. Prog. Phys.}

\newcommand{\JjphysA}{J. Phys. A: Math. Theor.}
\newcommand{\JjphysB}{J. Phys. B: At. Mol. Opt. Phys.}

\newcommand{\Jadvphys}{Adv. Phys.}

\newcommand{\Jannphys}{Ann. Phys. (NY)}

\newcommand{\JAnnualRevCondMat}{Annual Rev. Cond. Mat. Phys.}

\newcommand{\Jadvatmoloptphys}{Adv. At. Mol. Opt. Phys.}

\newcommand{\Jprogthphys}{Prog. Theor. Phys.}

\newcommand{\JZphysB}{Z. Phys. B}

\newcommand{\JRevSciInstrum}{Rev. Sci. Instrum.}

\usepackage{amssymb}
\usepackage{amsmath}
\usepackage[english,french]{babel}

\newtheorem{e-proposition}[theorem]{Proposition}

\newtheorem{e-definition}[theorem]{Definition\rm}

\renewcommand{\theequation}{\arabic{equation}}
\usepackage{graphicx}

\def\og{\leavevmode\raise.3ex\hbox{$\scriptscriptstyle\langle\!\langle$~}}
\def\fg{\leavevmode\raise.3ex\hbox{~$\!\scriptscriptstyle\,\rangle\!\rangle$}}

\usepackage{color}

\newcommand{\kL}{k_{\textrm{\tiny L}}}
\newcommand{\lambdaL}{\lambda_{\textrm{\tiny L}}}
\newcommand{\omegaL}{\omega_{\textrm{\tiny L}}}
\newcommand{\Er}{E_{\textrm{r}}}

\newcommand{\omegaA}{\omega_{\textrm{\tiny A}}}
\newcommand{\omegaZero}{\omega_{0}}
\newcommand{\asc}{a_{\textrm{sc}}}

\newcommand{\EF}{E_{\textrm{\tiny F}}}
\newcommand{\TF}{T_{\textrm{\tiny F}}}
\newcommand{\mF}{m_{F}}

\newcommand{\kB}{k_{\textrm{\tiny B}}}

\newcommand{\tunnel}{t}
\newcommand{\SupExch}{J}

\begin{document}
\centerline{Quantum simulation / Simulation quantique}
\begin{frontmatter}


\selectlanguage{english}
\title{Quantum simulation of the Hubbard model with ultracold fermions in optical lattices}
\selectlanguage{english}

\author[authorlabel1]{Leticia Tarruell},
\ead{leticia.tarruell@icfo.eu}
\address[authorlabel1]{ICFO-Institut de Ciencies Fotoniques, The Barcelona Institute of Science and Technology, 08860 Castelldefels (Barcelona), Spain}

\author[authorlabel2]{Laurent Sanchez-Palencia}
\ead{lsp@cpht.polytechnique.fr}
\address[authorlabel2]{CPHT, Ecole Polytechnique, CNRS, Universit\'e Paris-Saclay, Route de Saclay, 91128 Palaiseau cedex, France}

\begin{abstract}
Ultracold atomic gases provide a fantastic platform to implement quantum simulators and investigate a variety of models initially introduced in condensed matter physics or other areas. One of the most promising applications of quantum simulation is the study of strongly correlated Fermi gases, for which exact theoretical results are not always possible with state-of-the-art approaches. Here, we review recent progress of the quantum simulation of the emblematic Fermi-Hubbard model with ultracold atoms.
After introducing the Fermi-Hubbard model in the context of condensed matter, its implementation in ultracold atom systems, and its phase diagram, we review landmark experimental achievements, from the early observation of the onset of quantum degeneracy and superfluidity to demonstration of the Mott insulator regime and the emergence of long-range anti-ferromagnetic order.
We conclude by discussing future challenges, including the possible observation of high-Tc superconductivity, transport properties, and the interplay of strong correlations and disorder or topology.
\\
\noindent{\it To cite this article: L.~Tarruell and L.~Sanchez-Palencia, C. R. Physique X (2018).}

\vskip 0.5\baselineskip

\selectlanguage{french}
\noindent{\bf R\'esum\'e}
\vskip 0.5\baselineskip
\noindent
{\bf Simulation quantique du mod\`ele de Hubbard avec des fermions ultrafroids dans des r\'eseaux optiques. }
Les gaz atomiques ultrafroids offrent un excellente plateforme pour r\'ealiser des simulateurs quantiques et \'etudier une grande diversit\'e de mod\`eles introduits initialement en physique de la mati\`ere condens\'ee ou d'autres domaines. L'une des applications les plus prometteuses de la simulation quantique est l'\'etude des gaz de Fermi fortement corr\'el\'es, pour lesquels des r\'esultats th\'eoriques exacts ne sont pas toujours disponibles.
Nous  pr\'esentons ici une revue des progr\`es r\'ealis\'es r\'ecemment sur la simulation quantique de l'embl\'ematique mod\`ele de Fermi-Hubbard avec des atomes ultrafroids.
Apr\`es avoir pr\'esent\'e le mod\`ele de Fermi-Hubbard dans le contexte de la mati\`ere condens\'ee, sa r\'ealisation avec des atomes ultrafroids et son diagramme de phase, nous pr\'esentons les r\'ealisations exp\'erimentales les plus marquantes, de l'observation initiale de l'apparition de la d\'eg\'en\'erescence quantique et de la superfluidit\'e fermioniques \`a la mise en \'evidence du r\'egime de l'isolant de Mott et de l'\'emergence d'un ordre anti-ferromagn\'etique \`a longue port\'ee.
Nous concluons par une discussion des d\'efis futurs, dont la possibilit\'e d'observer la supraconductivit\'e \`a haute temp\'erature, les propri\'et\'es de transport et la comp\'etition de fortes corr\'elations et du d\'esordre ou de la topologie.
\\
\noindent{\it Pour citer cet article~: L.~Tarruell and L.~Sanchez-Palencia, C. R. Physique X (2018).}

 \vskip 0.5\baselineskip
\noindent{\small{\it Mots-cl\'es~:} Gaz de Fermi~; R\'eseaux optiques~; Transition de Mott~; Magn\'etisme quantique}}
\end{abstract}
\end{frontmatter}

\selectlanguage{english}

\tableofcontents

\section{Introduction}
In the last twenty years, ultracold quantum gases have emerged as a flexible platform for the study of a plethora of basic phenomena relevant to other fields, in particular to condensed-matter physics.
Following the development of a variety of control tools in atomic and laser physics, ultracold atoms now allow for a nearly perfect realization of model Hamiltonians originally introduced in the context of solid-state physics.
A major asset of these systems is that most of the physical parameters, such as the spin, the interactions, the potential landscape, as well as the dimension, can be controlled precisely and varied in wide ranges.
Furthermore, ultracold atoms give access to novel observables~--~down to the single-atom level~--~ and extreme parameter regimes,
that are complementary to those accessible in solid-state systems.
They can thus be seen as an ideal realization of Feynman's vision of an analogue quantum simulator~\cite{feynman1982,lloyd1996}.
A number of benchmark demonstrations have already been reported first with bosonic gases and then with fermionic ones,
for recent reviews see Refs.~\cite{lewenstein2007,bloch2008,lewenstein2012,NaturePhysicsInsight2012bloch}.

Quantum degenerate gases of fermionic isotopes of atoms are particularly well suited to the simulation of solid-state systems, because they obey Fermi-Dirac statistics, which allows them to mimic directly the behaviour of the electrons in a solid. Since the first realization of a degenerate gas of ultracold fermionic atoms in 1999~\cite{demarco1999}, several fermionic atomic species have been cooled down to quantum degeneracy and are currently available for these studies. They include
alkali atoms (potassium $^{40}$K~\cite{demarco1999} and lithium $^{6}$Li~\cite{truscott2001,schreck2001}),
metastable atoms (helium $^3$He$^*$~\cite{mcnamara2006}),
two-electron atoms (ytterbium $^{173}$Yb~\cite{fukuhara2007} and strontium $^{87}$Sr~\cite{desalvo2010,tey2010}),
as well as long-range interacting dipolar atoms (dysprosium $^{161}$Dy~\cite{lu2012}, erbium  $^{167}$Er~\cite{aikawa2014}, and chromium $^{53}$Cr~\cite{naylor2015}). In these systems, the role of the electronic $1/2$-spin state is played by two distinct atomic internal states.
Furthermore, several of these atomic species allow for the precise tuning of the strength and sign of the interactions between the atoms \emph{via} a homogeneous magnetic field, using so-called Feshbach resonance techniques~\cite{chin2010}. This gives access to regimes of strong interactions, which are the hallmark of strongly correlated electronic materials where quantum many-body effects are prominent. A theoretical description of such systems remains a major challenge to many-body physics and quantum simulation is most desirable.

The first quantum simulation studies with ultracold fermionic atoms focused on the study of superfluidity in trapped two-component Fermi gases with strong interactions~\cite{zwerger2011}, for instance the BEC-BCS crossover~\cite{greiner2003,jochim2003,zwierlein2003,chin2004,regal2004,zwierlein2004,bourdel2004} and various thermodynamic properties~\cite{nascimbene2009,nascimbene2010,navon2010,houcke2012,ku2012}.
Moreover, a more complete analogy to solid-state systems can be realized in the presence of an optical lattice. The latter is obtained from the interference pattern of an ensemble of laser beams and generates a spatially periodic potential for the atoms. It hence provides an almost perfect realization of the crystalline lattice of a real solid that is nearly free of defects and phonons~\cite{lewenstein2007,lewenstein2012,grynberg2001,bloch2005}.
Depending on the exact geometrical arrangement of the laser beams, a broad range of lattice geometries is possible,
from simple square or cubic lattices~\cite{greiner2002} to more complex dimerized, triangular, hexagonal, Kagome or Lieb geometries \cite{sebby2006,anderlini2007,folling2007,becker2010,soltan2011,tarruell2012,jo2012,taie2015}. For a review see Ref.~\cite{windpassinger2013}.

In this paper, we review recent progress on the quantum simulation of the Hubbard model with ultracold Fermi gases in optical lattices and focus on the simple square and cubic geometries. The Hubbard Hamiltonian is the pivotal model for strongly correlated electronic materials.
It is exactly solvable only in special cases, for instance in one dimension~\cite{lieb1968,lieb2003} and in the limit of infinitely connected lattices~\cite{georges1992}.
In all the other cases, including the two- and three-dimensional lattices that naturally describe real materials, it is particularly appealing for quantum simulation. This approach complements existing powerful theoretical and numerical approaches, such as quantum Monte Carlo (QMC)~\cite{suzuki1993} and dynamical mean field theory (DMFT)~\cite{georges1996}, to check their predictions, provide new insights in regimes not accessible to these techniques, and possibly guide the development of new theoretical approaches.
At low energy, the interplay between the density and spin degrees of freedom gives rise to a very rich physics, encompassing distinct transport regimes, and magnetically ordered phases.
While the exact phase diagram of the Hubbard model is not rigorously known,
simple arguments provide a general picture at half filling,
that is when the lowest Bloch band of each spin state is half-filled.
For repulsive interactions, it includes Mott-insulating and magnetically ordered phases as predicted theoretically and fairly assessed numerically.
In the presence of doping, however, the situation is completely open but several arguments suggest that the two-dimensional Hubbard model may explain high-$T_\textrm{c}$ superconductivity~\cite{anderson1987}. Yet, a complete theoretical understanding of the model is still lacking, owing to the difficulty of its numerical simulation~\cite{micnas1990}. It is thus expected that quantum simulation offers a new perspective over the problem, and guides theoretical approaches~\cite{hofstetter2002}.

The present review aims at providing a sufficiently didactic, although non-technical, introduction to the physics of the Fermi-Hubbard model and its simulation with ultracold atoms. It complements other reviews that appeared recently, see for instance Refs.~\cite{esslinger2010,hofstetter2018}.
It is our aim that the review is accessible and useful to both the condensed-matter (CM) and atomic, molecular, and optical (AMO) communities.
For this reason, Sec.~\ref{sec:FHm} combines parts that may be more useful to one community than to the other.
For instance, the construction of the Hubbard model in solid-state physics (Sec.~\ref{sec:FHmCM}) and the discussion of the phase diagram~(\ref{sec:FHmPhasDiag}) may be more useful to the AMO community but may appear standard to specialists of strongly correlated materials.
In turn, the derivation of the Hubbard model from first principles and its realization in the context of ultracold atoms~(\ref{sec:FHmUA}), which includes a discussion of basic control tools in these systems, may be more useful to the CM community.
The remainder of the review bridges the gap and illustrates how the powerful control tools and the variety of probes at the atomic scale available in AMO physics are providing new insights in classic questions of CM physics.

More precisely, the review is organized as follows.
In Sec.~\ref{sec:FHm}, we provide a comprehensive introduction to the Fermi-Hubbard model and its basic phase diagram.
We introduce the model in the context of condensed-matter physics (Sec.~\ref{sec:FHmCM}) and describe its realization in ultracold-atom systems (Sec.~\ref{sec:FHmUA}).
The phase diagram at half filling is then discussed exploiting the spin-charge separation, which enlightens the distinction between the transport properties and the emergence of quantum magnetism (Sec.~\ref{sec:FHmPhasDiag}).
In Sec.~\ref{sec:FirstSimulations}, we review early quantum simulation experiments of the Fermi-Hubbard
and show how the special features of ultracold atoms allow for the direct observation of basic properties of lattice Fermi gases, such as fermionic quantum degeneracy (Sec.~\ref {sec:NonInteractingFermi}) and superfluidity (Sec.~\ref{sec:Superfluidity}).
The next sections are devoted to a series of experiments that have successfully demonstrated a number of hallmark features of the strongly interacting regime, from the observation of the Fermi Mott insulator to the emergence of quantum anti-ferromagnetism.
Section~\ref{sec:observMott} focuses on the experimental characterization of the Mott insulator state, using bulk measurements (Sec.~\ref{sec:MIbulk}) and spatially resolved probes (Sec.~\ref{sec:spatiallyresolved}), with special emphasis on the two-dimensional Hubbard model (Sec.~\ref{sec:2D}).
Section~\ref{sec:AForder} focuses on recent experimental studies of the magnetic properties of the Hubbard model,
from the onset of short-range anti-ferromagnetic order (Sec.~\ref{sec:spinorder}) to the demonstration of long-range anti-ferromagnetism (Sec.~\ref{sec:AFMorder}).
Finally, the outlook in Sec.~\ref{sec:outlook} discusses complementary active research lines and future challenges for the quantum simulation of strongly correlated Fermi gases.

\section{The Fermi-Hubbard model and its many-body phase diagram}\label{sec:FHm}

In this section, we review basic knowledge of the Fermi-Hubbard model and its phase diagram. We first discuss its relevance in the context of condensed-matter physics (Sec.~\ref{sec:FHmCM}) and its simulation with ultracold atoms (Sec.~\ref{sec:FHmUA}). We then discuss the  phase diagram, focusing on the half-filling case (Sec.~\ref{sec:FHmPhasDiag}).

\subsection{The Fermi-Hubbard Hamiltonian as a toy model for strongly correlated electrons in solids}
\label{sec:FHmCM}

\begin{figure}
\begin{center}
\includegraphics[width=0.95\hsize]{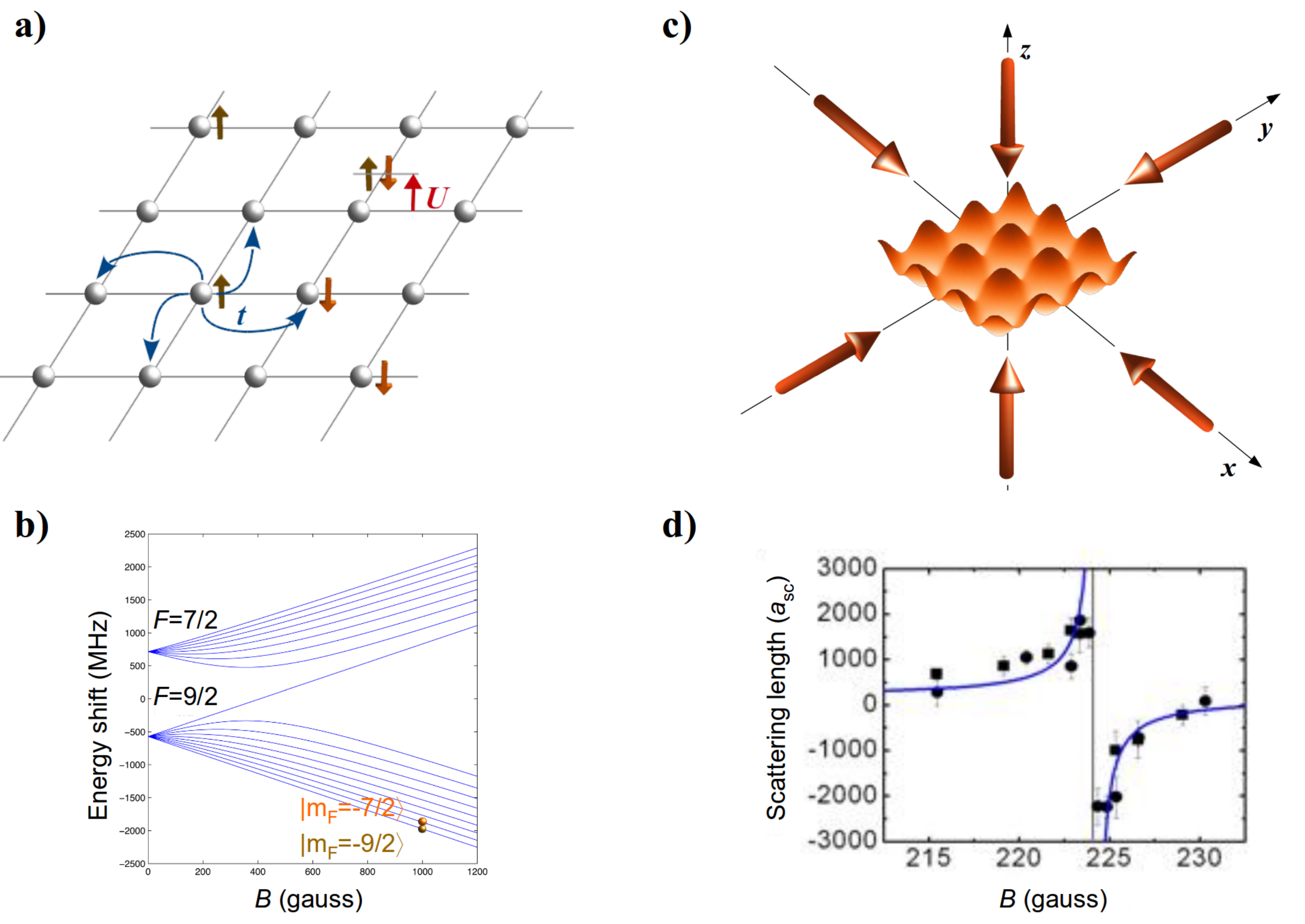}
\caption{\label{fig:FHmCMUA}
Fermi-Hubbard model and its realization with ultracold atoms.
(a)~Schematic representation of the Fermi-Hubbard model. A solid-state crystal realizes a periodic array of ions (gray blue spheres, here shown in two dimensions). The electrons of the conduction band, either in the spin $\uparrow$ (brown arrow) of spin $\downarrow$ (orange arrow) state, are tightly bound to the ions. They can tunnel to the neighboring sites with the hopping amplitude $\tunnel$. When two electrons, in opposite spin states, are in the same site, they interact with the interaction energy $U$.
(b)~Internal-state structure of the $^{40}$K atom in the $F=9/2$ (lower manifold) and $F=7/2$ (higher manifold) hyperfine level. The atom is placed in a magnetic field, which Zeeman shifts the various states.
In general two states, for instance $\vert F=9/2,\mF=-9/2\rangle$ and $\vert F=9/2,\mF=-7/2\rangle$, encode the two spin components of the Fermi-Hubbard model.
Ancilla states in the $F=7/2$ manifold may be used as virtual states to laser manipulate the state of the system.
Panel extracted and adapted from Ref.~\cite{leblanc2011phd}.
(c)~Three-dimensional optical lattice as realized from the interference pattern of three orthogonal pairs of counter-propagating laser beams along the main axes $x$, $y$, and $z$. This creates a simple cubic periodic array of lattice sites (shown in two dimensions) which mimic the ion sites of the Hubbard model.
(d)~Feshbach control of interactions. The figure shows the scattering length $\asc$ versus the magnetic field for collisions between $^{40}$K atoms in the $\vert F=9/2,\mF=-9/2\rangle$ and $\vert F=9/2,\mF=-5/2\rangle$ states.
The strong variation of $\asc$, with a sign-changing divergence (Feshbach resonance), permits to control the strength of the interactions, including attractive ($\asc<0$), repulsive ($\asc>0$), and vanishingly small ($\asc \simeq 0$) interactions.
Panel extracted from Ref.~\cite{regal2003}.
}
\end{center}
\end{figure}

The Hubbard model describes the dynamics of correlated quantum particles in a discrete lattice. It was originally introduced in the context of solid-state physics to understand the electronic dynamics and the magnetic properties of strongly correlated materials~\cite{gutzwiller1963,hubbard1963,gutzwiller1964}. In such systems, the constitutive crystal ions form a spatially periodic lattice, see Fig.~\ref{fig:FHmCMUA}(a).
Each ion occupies a lattice site $j$ and host electrons from the conduction band. The electron dynamics is then governed by two elementary processes.
Firstly, the electrons of the conduction band can tunnel from an ion of the crystal lattice to another ion, say from site $j$ to site $\ell$. The tunnelling probability results from the spatial overlap of the electronic orbital wavefunctions and decays fast with the distance. One may retain only the dominant terms corresponding to hopping to the nearest-neighbor crystal ions. This yields the tight-binding (single-band) kinetic term $\hat{H}_0=-\tunnel\sum_{\langle j,\ell\rangle,\sigma} \left( \hat{c}^{\dagger}_{j,\sigma} \hat{c}_{\ell,\sigma}+\mathrm{h.c.}\right)$, where $\tunnel$ is the tunnelling energy,
$\hat{c}^{\dagger}_j$ and $\hat{c}_j$ denote, respectively, the creation and annihilation operators of an electron of spin $\sigma\in\{\uparrow,\downarrow\}$ on the lattice site $j$, and the notation $\langle j,\ell\rangle$ restricts the summation to all pairs of nearest neighbor sites $j$ and $\ell$.
Secondly, the electrons are coupled via the Coulomb interaction. This yields two-body repulsive interactions that decay with the distance. In first approximation and assuming significant screening of the long-range Coulomb potential, one may retain only the dominant term corresponding to the interaction energy $U$ between two electrons on the same lattice site. Owing to the Pauli principle, only two electrons of opposite spins may occupy the same site and the interaction term reduces to $\hat{H}_1= U\sum_j\hat{n}_{j,\uparrow}\hat{n}_{j,\downarrow}$, where $\hat{n}_{j,\sigma}=\hat{c}^{\dagger}_{j,\sigma} \hat{c}_{j,\sigma}$ denotes the number operator for electrons of spin $\sigma$ in the lattice site $j$.
The resulting Hubbard Hamiltonian,
\begin{equation}
\hat{H}=-\tunnel\sum_{\langle j,\ell\rangle,\sigma} \left( \hat{c}^{\dagger}_{j,\sigma} \hat{c}_{\ell,\sigma}+\mathrm{h.c.}\right)+U\sum_j\hat{n}_{j,\uparrow}\hat{n}_{j,\downarrow},\label{FH-Hamiltonian}
\end{equation}
is the basic model to understand the transport and the magnetism of transition metals.
It is generally considered an excellent model for narrow-band materials where a single atomic orbital per lattice site is sufficient~\cite{mahan2000}.
More generally it constitutes a universal model for describing the physics of strongly correlated materials~\cite{Sachdev2001}.

In the context of solid-state physics, the Hubbard Hamiltonian~(\ref{FH-Hamiltonian}) is essentially a heuristic model, and the parameters $\tunnel$ and $U$ may be chosen a posteriori by comparison to intensive numerical calculations or experimental data.
The ab initio determination of the Hubbard parameters $\tunnel$ and $U$ is a formidable task, and it is still the object of intensive research. The main difficulty is that the electrons of the conduction band participate not only to the electronic conduction but also to the stability of the lattice structure. The derivation of the Hubbard model thus requires in principle to solve the many-body Schr\"odinger equation for both the electrons and the ions.
Moreover, the Hubbard model ignores several effects, which may significantly affect the electronic properties. For instance, the zero-point crystal structure is subjected to dynamical distortion induced by quantum and thermal fluctuations. These are described by phonon modes, which interact with the electrons. This significantly affects the transport properties of nearly any material. It can also have strong effects including the creation of phonon-electron pairs (polarons), or an effective attraction between the electrons, responsible for the formation of Cooper pairs and Bardeen-Cooper-Schrieffer (BCS) superconductivity. In first approximation, many of these effects may be accounted for by the renormalization of the Hubbard parameters.
For instance, BCS superconductivity may be addressed by assigning a negative value to the interaction energy $U$.
Other effects require to include beyond-Hubbard model terms.
While the reduction of the tunneling to the nearest neighbors is well justified in narrow-band materials, the reduction of the interaction to a local term is more delicate since it originates from a long-range term. In fact interactions to distant lattice sites may not be non-negligible in real materials.
For instance, the determination of charge fluctuations requires off-diagonal terms~\cite{mahan2000}.
In contrast, the Hubbard model is often sufficient to describe spin fluctuations.

\subsection{Realization of the Fermi-Hubbard Hamiltonian with ultracold atoms in optical lattices}
\label{sec:FHmUA}

We now turn to ultracold atomic systems. A major asset of the latter is that the Hubbard model can be derived with controlled accuracy from first principles. Here we briefly outline the derivation. We work along the lines of Ref.~\cite{Jaksch1998} where the Hubbard model was first derived for ultracold bosons in optical lattices, and adapt the discussion to fermions. For more details, the interested reader may refer to other reviews~\cite{jaksch2005,lewenstein2007,bloch2008}.

\paragraph*{Many-body Hamiltonian~--}
Ultracold-atom systems are dilute atomic gases that are cooled down using a sequence of laser and evaporative cooling techniques~\cite{chu1998,cct1998,phillips1998,ketterle1996}. Once cooled down to quantum degeneracy, typically in the $1-500$~nK range, they are spatially confined in optical traps~\cite{ketterle1999,dalfovo1999,ketterle2008,giorgini2008}.
The low-energy physics of a two-component ultracold Fermi gas as considered here is then governed by the many-body Hamiltonian
\begin{equation}\label{Hamiltonian}
\hat{H} =
\sum_{\sigma=\uparrow,\downarrow} \int d\vec{r}\, \hat{\Psi}_\sigma^\dagger(\vec{r}) \left[ \frac{-\hbar^2 \vec{\nabla}^2}{2m} + V(\vec{r})\right] \hat{\Psi}_\sigma(\vec{r})
+ g \int d\vec{r}\,  \hat{\Psi}_\uparrow^\dagger(\vec{r}) \hat{\Psi}_\downarrow^\dagger(\vec{r}) \hat{\Psi}_\downarrow(\vec{r}) \hat{\Psi}_\uparrow(\vec{r}),
\end{equation}
with $m$ the atomic mass and $\hbar$ the reduced Planck constant. The quantity $\hat{\Psi}_\sigma(\vec{r})$ is the continuous-space field operator for the spin component $\sigma$ at position $\vec{r}$.
The first term in Eq.~(\ref{Hamiltonian}) accounts for the kinetic energy and the interaction with some external potential $V(\vec{r})$. The latter is assumed to be independent of the spin component.
It is realized by exploiting the interaction of the neutral atoms with far from resonance laser fields (see below).
The second term describes two-body contact interactions between fermions in opposite spin components, with the coupling constant $g$.

The two components $\sigma$ are encoded in two different internal atomic states.
In general, the atoms are in their electronic ground state. The latter is a hyperfine manifold of degeneracy $F$, see an example on Fig.~\ref{fig:FHmCMUA}(b). The degeneracy of the hyperfine states is lifted by the Zeeman effect in a homogeneous magnetic field.
The atoms are then prepared in a particular hyperfine state, for instance $\vert F=9/2, \mF=-9/2\rangle$ as in the figure. By tuning radio-frequency pulse, one then couples the latter to another state, for instance $\vert F=9/2, \mF=-7/2\rangle$, and forms a mixture of atoms in the two corresponding Zeeman states with a controlled balance. The states are labelled by the pseudo-spin variable $\sigma \in \{\uparrow, \downarrow\}$.

In the atomic gas, the electrically neutral atoms interact via two-body van der Waals interactions
with a short-distance hard-core cut-off at the atomic-radius distance scale. For temperatures typically below the millikelvin range, the scattering states with a non-zero angular momentum are energetically suppressed owing to the scattering centrifugal barrier. Then, only the s-wave scattering harmonics is relevant. Being rotation invariant, it may be rigorously modeled by a pure contact interaction potential. The latter is characterized by a single parameter, known as the scattering length $\asc$. The coupling constant in Eq.~(\ref{Hamiltonian}) is related to the latter via the formula $g = 4\pi\hbar^2\asc/m$. Since the s-wave harmonic is symmetric, the Pauli principle acts on the spin component and only two fermions in opposite spin states can interact, as described in Eq.~(\ref{Hamiltonian}).

The Hamiltonian~(\ref{Hamiltonian}) is in most cases sufficient to accurately describe ultracold Fermi gases. However, the high flexibility of ultracold atoms also allows to realize more exotic situations with similar techniques. For instance, the external potential can be designed to be spin-dependent by shifting the Zeeman sublevels using AC Stark or Zeeman effect, and using two laser fields, each one being near resonant with one of the states~\cite{liu2004}.
Fermi gases with more than two spin components can also be realized by straightforward extension of the method outlined above. For instance, Fermi gases with $SU(\kappa)$ symmetry and an adjustable value of the spin degeneracy $\kappa$ up to $6$ have been demonstrated in Ref.~\cite{taie2012,pagano2014,hofrichter2016}. Finally, long-range, dipolar interactions can be implemented using atomic species with large magnetic moments~\cite{lahaye2009} or polar molecules~\cite{micheli2006}. We, however, disregard such cases in the remainder of this review.

\paragraph*{Dipole traps and optical lattices~--}
The external potential $V(\vec{r})$ is created with controlled electromagnetic fields, most often with laser light.
Optical design and interference of several laser beams permit to create a wide variety of controlled potentials, including
deep traps and narrow barriers~\cite{grimm2000},
periodic potentials in various crystalline configurations~\cite{grynberg2001,bloch2005,greiner2002,sebby2006,anderlini2007,folling2007,becker2010,soltan2011,tarruell2012,jo2012,taie2015,windpassinger2013},
as well as quasi-periodic lattices~\cite{guidoni1999,modugno2010,lsp2005}
and disordered potentials~\cite{fallani2008,lsp2010,shapiro2012}.

The optical control of the potential exploits the dipole force induced by the inhomogeneous AC Stark effect. One uses an oscillating laser field quasi-resonant with an internal atomic transition. The laser-atom interaction creates an oscillating atomic dipole parallel and proportional to the local electric field of the laser. The latter interacts back to the same electric field via the dipole interaction, hence creating the optical potential
\begin{equation}\label{DipolePotential}
V(\vec{r}) = d_0^2 \mathcal{E}(\vec{r})^2 / 4 \hbar \delta,
\end{equation}
where $d_0$ is the matrix element of the atomic dipole, $\mathcal{E}(\vec{r})$ is the amplitude of the electric field, and $\delta=\omegaL-\omegaA$ is the detuning between the laser ($\omegaL$) and atomic ($\omegaA$) angular frequencies.
The dipole potential~(\ref{DipolePotential}) is proportional to the laser intensity $I(\vec{r}) \propto \mathcal{E}(\vec{r})^2$ and inversely proportional to the laser detuning $\delta$, which can both be tuned experimentally. This permits to control the sign (via $\delta$) and the amplitude (via $I$ and $\vert\delta\vert$) of the dipole potential. For instance, a focused laser beam creates a local maximum of the intensity. For $\delta>0$ (so called \textit{blue detuning}), it creates a repulsive potential near of maximum of intensity. For $\delta<0$ (\textit{red detuning}), it creates attractive potential where the atoms can be trapped.

In this review, we focus on \textit{periodic lattices}. The simplest case is obtained by shinning two counter-propagating laser beams of identical frequencies, say along the $x$ axis, see Fig.~\ref{fig:FHmCMUA}(c). They create a standing wave, $\mathcal{E}(x) \propto \sin(\kL x)$, and thus the periodic potential $V_x(x) = V_0^x \sin^2(\kL x)$. Here $\kL = 2\pi/\lambdaL$ and $\lambdaL$ are, respectively, the laser wavevector and wavelength, and the spatial origin is chosen adequately. Using three pairs of counter-propagating laser beams along the $x$, $y$, and $z$ directions respectively, one then creates the periodic potential
\begin{equation}\label{PeriodicPotential}
V(\vec{r}) = V_0 \Big[\sin^2(\kL x) + \sin^2(\kL y) + \sin^2(\kL z) \Big].
\end{equation}
In practice, the realization of such a potential requires the suppression of interference between the three pairs of laser beams creating the sine potential in each spatial direction. This is achieved by using mutually orthogonal polarizations or laser frequencies that are slightly detuned by values exceeding the inverse of the typical motional timescale so that the beating between the laser pairs averages out.
The slight differences it induces in the spatial periods of the pairs is insignificant.

The simple configuration we have considered above forms a spatially periodic lattice of spacing $a = \pi/\kL=\lambdaL/2$ in all the main spatial directions $x$, $y$, and $z$. This is a simple cubic three-dimensional Bravais lattice~\cite{ashcroft1976}.
A variety of extensions can be realized using more complicated laser configurations. For instance, extensions to one and two dimensions are straightforward by using the corresponding number of counter-propagating pairs. Trapping along the complementary direction(s) is then provided by the finite transverse extension of the laser beams, which we have disregarded so far. More complicated Bravais lattice structures can also be designed using the interference pattern of more complicated laser arrangements~\cite{grynberg2001,windpassinger2013}.

\begin{figure}
\begin{center}
\includegraphics[width=0.5\hsize]{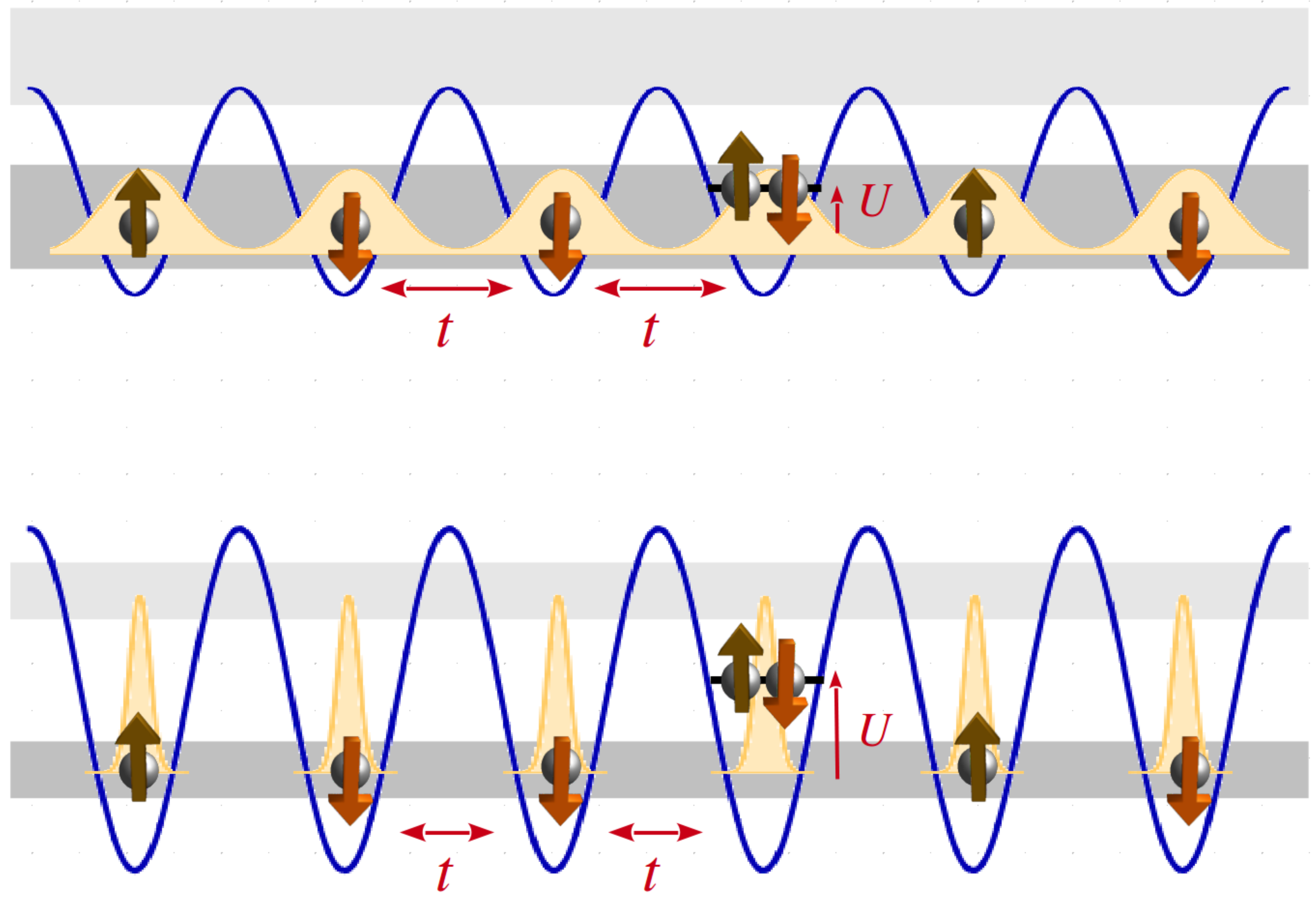}
\caption{\label{fig:bands}
Schematic representation of the Bloch-band structure and the basic elements of the Fermi-Hubbard model.
In a periodic potential (solid blue lines), the single-particle eigenstates are organized in energy bands (shaded gray zones). The corresponding eigenstates are delocalized Bloch states. Each Bloch state is a superposition of Wannier states (yellow curves). The Wannier states are all localized in a different lattice site and are coupled by quantum tunneling, with the tunneling energy $t$. In the Hubbard model, two fermions in spin up (brown) and down (orange) states interact with the interaction energy $U$.
When the amplitude of the periodic potential increases (from top to bottom),
the Bloch band shrinks in energy and Wannier states get narrower. Then, the tunneling energy decreases and the interaction energy increases.
}
\end{center}
\end{figure}

\paragraph*{Derivation of the Fermi-Hubbard model~--}
To derive the Fermi-Hubbard model, it is worth starting with the non-interacting case, corresponding to the first term in the right-hand-side of Eq.~(\ref{Hamiltonian}). The latter describes single non-relativistic particles in a periodic potential, that is the paradigmatic model of the Bloch theory of solids~\cite{ashcroft1976}. It follows from the Bloch theorem that the single-particle eigenstates are extended Bloch states arranged in energy bands separated by finite band gaps, see Fig.~\ref{fig:bands}. When the typical energies of the Fermi gas, namely the chemical potential, the temperature, and, for interacting particles, the two-particle interaction energy, are much lower than the gap to the first excited band, we may restrict ourselves to the lowest energy band.
The latter is spanned by the corresponding Bloch basis,
or alternatively by the Wannier basis~\cite{ashcroft1976}.
The Wannier states are constructed as linear combinations of the Bloch states so as to create a set of single-particle wavefunctions $w_j(\vec{r})$ localized in the minima $j$ of the periodic potential, see Fig.~\ref{fig:bands}.
Since the potential is spatially periodic, all the Wannier states are identical up to a translation from one site to another, $w_j(\vec{r})=w_0(\vec{r}-\vec{r}_j+\vec{r}_0)$. Note that the Wannier functions inherit the spin independence of the single-particle part of the Hamiltonian~(\ref{Hamiltonian}).

Working within the lowest-energy band, the field operator may be decomposed into the corresponding Wannier basis, $\hat{\Psi}_\sigma(\vec{r})=\sum_j w_j(\vec{r})\hat{c}_{j,\sigma}$, where $\hat{c}_{j,\sigma}$ annihilates a fermion of spin $\sigma$ into the Wannier state at site $j$. Inserting this decomposition into the Hamiltonian~(\ref{Hamiltonian}) and retaining the dominant terms yields the Fermi-Hubbard model~(\ref{FH-Hamiltonian}). The first term in Eq.~(\ref{Hamiltonian}) yields the single-particle Hamiltonian $\hat{H}_1 = - \sum_{j,\ell, \sigma} \tunnel_{j,\ell} \hat{c}_{j,\sigma}^\dagger\hat{c}_{\ell,\sigma}$, where
$\tunnel_{j,\ell}=-\int d\vec{r}\, w_j^*(\vec{r}) \Big[ {-\hbar^2 \vec{\nabla}^2}/{2m} + V(\vec{r})\Big] w_\ell(\vec{r})$
and the sum runs over all pairs of lattice sites, $(j,\ell)$. Since the Wannier functions are localized in the vicinity of the lattice sites, the quantity $\tunnel_{j,\ell}$ decays fast with the distance $\vert\vec{r_\ell-\vec{r}_j}\vert$. Apart from the terms corresponding to $j=\ell$, which only produce a local potential (see below), the dominant terms correspond to adjacent lattice sites, i.e.\ $j$ and $\ell$ such that $\vert\vec{r_\ell-\vec{r}_j}\vert=a$.
It yields the first term of the Fermi-Hubbard Hamiltonian~(\ref{FH-Hamiltonian}), with $\tunnel=\tunnel_{j,\ell}$ for $j$ and $\ell$ corresponding to nearest neighbor lattice sites\footnote{In the present notations, each pair is counted only once. In other words, the pairs $\langle j, \ell \rangle$ and $\langle \ell, j \rangle$ are not distinguished.}.
For the separable sine potential~(\ref{PeriodicPotential}), it can be
found by solving the Mathieu equation~\cite{campbell1955}. For $V_0 \gg E_\textrm{r}$, where $\Er = \hbar^2 \kL^2/2m$ is the recoil energy, one finds
\begin{equation}\label{Tunneling}
\frac{\tunnel}{E_\textrm{r}} \simeq \frac{4}{\sqrt{\pi}} \left(\frac{V_0}{E_\textrm{r}}\right)^{3/4} \exp\left(-2\sqrt{V_0/E_\textrm{r}}\right).
\end{equation}
The interaction term is found similarly by inserting the decomposition $\hat{\Psi}_\sigma(\vec{r})=\sum_j w_j(\vec{r})\hat{c}_{j,\sigma}$ into the second term in Eq.~(\ref{Hamiltonian}). Retaining only the dominant term, which now amounts to take all the Wannier functions in the same lattice site, one finds the second term of the Fermi-Hubbard Hamiltonian~(\ref{FH-Hamiltonian}), with $U = g \int d\vec{r}\, \vert w_j(\vec{r}) \vert^4$. Note that, owing to the discrete translation invariance, this term is independent of the lattice site $j$.
The calculation of the two-fermion interaction energy is simpler than for the tunneling energy. Here an accurate value may be found using a harmonic expansion of the lattice potential,
$V(\vec{r}) \simeq V_0 \kL^2 \vec{r}^2 = m\omegaZero^2 \vec{r}^2/2$ with $\omegaZero = 2\sqrt{V_0 E_\textrm{r}}/\hbar$. The Wannier function of the lowest energy band is then assimilated to the ground state of the harmonic oscillator, $w_0(\vec{r}) \simeq (\pi\ell_0^2)^{-3/4}\exp\big(\vec{r}^2/2\ell_0^2\big)$. Inserting this expression into the formula for $U$, one then finds
\begin{equation}\label{InteractionU}
\frac{U}{E_\textrm{r}} \simeq \sqrt{8/\pi} \ \kL \asc \, \left(\frac{V_0}{E_\textrm{r}}\right)^{3/4}.
\end{equation}

\paragraph*{Experimental control of the Hubbard parameters~--}
As discussed above, the Fermi-Hubbard model~(\ref{FH-Hamiltonian}) is derived from first principles with accurate values of the relevant parameters $\tunnel$ and $U$.
The only used approximations are the restrictions to (i)~the lowest Bloch band, (ii)~nearest-neighbor tunneling, and (iii)~on-site interactions.
While the single-band approximation~(i) is valid as long as the interaction strength $U$, the thermal energy $\kB T$, and the chemical potential $\mu$ are smaller than the energy of the first excited band, the narrow-band approximations~(ii) and (iii) are valid in deep enough lattice potentials. These conditions are fulfilled by the vast majority of experiments with ultracold atoms. Moreover, the use of external laser fields to create the artificial crystal permits a realization of the Fermi-Hubbard that is almost perfect. In contrast to solid-state systems, the periodic structure is here independent of charge carriers and does not need be determined self-consistently with the density of fermions. In addition, the stabilization of the laser beams strongly suppresses the vibration of the lattice and the associated heating effects.

A unique asset of the ultracold-atom realization of the Hubbard model is that it permits an accurate control of the parameters. For instance, the scaling of the tunneling ($\tunnel$) and interaction ($U$) energies with the amplitude of the lattice potential $V_0$ [see Eqs.~(\ref{Tunneling}) and (\ref{InteractionU})], permits to control the relevant parameter $\tunnel/U$ by adjusting the intensity of the laser beams, which is proportional to $V_0$, see Eq.~(\ref{DipolePotential}). Moreover, for a fixed amplitude of the lattice potential, the interaction strength $U$ can be independently controlled via the value of the scattering length $\asc$, using Feshbach resonance techniques~\cite{chin2010}.
In brief, a homogeneous magnetic field is applied to the atomic gas.
It produces a relative energy shift between scattering open channels and bound states.
When the two are close to resonance, they form long-lived quasi-bound states, which induce a resonance in the scattering cross section. The coupling constant then turns from negative to positive values with arbitrary absolute values, see Fig.~\ref{fig:FHmCMUA}(d).
Fine tuning of the external magnetic field then permits to choose both the value and the sign of $\asc$ within a broad range.
Feshbach resonance techniques have been developed for both ultracold Bose and Fermi gases. In practice, it is particularly suited for fermions because, in contrast to bosons, the three-body collisions are suppressed as a consequence of the Pauli exclusion rule~\cite{petrov2004b}.

Additional parameters can also be controlled.
For instance, in standard schemes the final cooling stage is performed by evaporation, which consists in a progressive removal of fractions of particles with the highest energies~\cite{ketterle1996}. This progressively lowers both the temperature $T$ and the total number of particles $N$, which can thus be controlled by stopping the evaporation stage appropriately.
In order to reach the lowest temperatures, more elaborate cooling schemes have been implemented in the last preparation stages. For instance, trap reshaping-induced redistribution of entropy has been used to lower the temperature below the N\'eel temperature and observe anti-ferromagnetic ordering in the Fermi-Hubbard model, see Ref.~\cite{mazurenko2017} and Sec.~\ref{sec:AFMorder}.
Finally, the balance between spin up and spin down components may be further controlled using a radio-frequency coupling pulse to transfer atoms from one component to the other.

\paragraph*{Inhomogeneous trapping~--}
So far, we have considered optical lattices as created by infinitely wide plane-wave laser fields. In practice, however, the laser beams have a transverse intensity distribution, typically Gaussian. As a result, the lattice amplitude $V_0$ varies smoothly in space, hence creating a trap (for red-detunned laser light) or an anti-trap (for blue detuning). In the latter case, additional trapping is necessary. It may be realized using an optical dipole potential, independent of the lattice. In both cases, it creates an inhomogeneous trapping term, which breaks the translation invariance. To avoid this issue, one may use box potentials~\cite{gaunt2013,chomaz2015,mukherjee2017,hueck2018}. In some cases, however, the inhomogeneous trap has major advantages.
For low energy, the trap may be described by the harmonic potential
$V_{\textrm{trap}}(\vec{r}) = m\omega_x^2 x^2/2 + m\omega_y^2 y^2/2 + m\omega_z^2 z^2/2$, where the $\omega_j$'s are the angular frequencies of the trap.
The Fermi-Hubbard Hamiltonian is then augmented by the additional term
\begin{equation}\label{TrappingTerm}
\hat{H}_\textrm{trap} \simeq V_j \hat{n}_j,
\end{equation}
where $V_j = \int d\vec{r}\ V_{\textrm{trap}}(\vec{r}) \vert w_j(\vec{r})\vert^2$.
For isotropic traps, one has $V_j \propto \vert j \vert^2$, where $\vert j \vert$ is the distance of the lattice site $j$ from the trap center, in units of the lattice spacing.
As we shall see, the latter may be used advantageously to control the lattice filling factor, that is the average number of particles per lattice site, near the trap center. In a local density approximation, it can be described as a spatially varying, effective, chemical potential $\mu(\vec{r}_j)=\mu-V_j$, where $\mu$ is the chemical potential in the trap center. Hence, a single realization of the experiment realizes a broad range of fillings.
More precisely, the density profile typically creates a wedding-cake structure~\cite{Jaksch1998}, which can be observed through position-resolved measurements (see below).

\subsection{Phase diagram of the Fermi-Hubbard model}
\label{sec:FHmPhasDiag}

The Fermi-Hubbard model displays a rich phase diagram, resulting from the interplay of interactions, quantum tunneling, Pauli exclusion, and the spin degrees of freedom.
It has been extensively studied in the context of theoretical condensed-matter physics. While an exact solution to the problem is unknown, except in one dimension~\cite{lieb1968,lieb2003}, a number of properties are well established, at least in some specific cases. Summarizing it in its full generality goes beyond the scope of this review. Here we focus on the case of half-filling, which is well understood.
It corresponds to the case where each lattice site is populated in average by one fermion and the spins are balanced, namely $\langle \hat{n}_{\uparrow}\rangle=\langle \hat{n}_{\downarrow}\rangle=1/2$.

\begin{figure}
\begin{center}
\includegraphics[width=0.95\hsize]{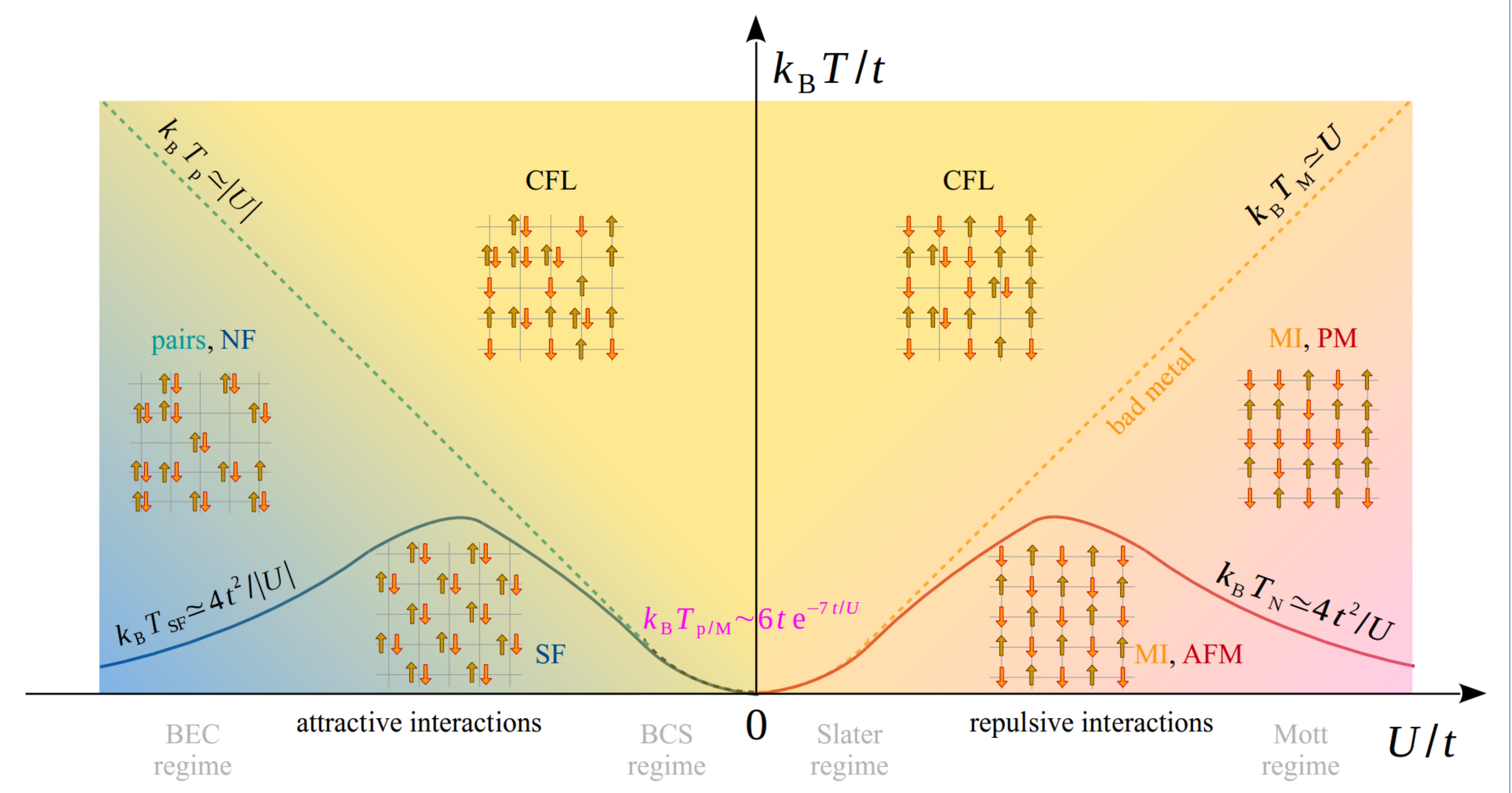}
\caption{\label{fig:FHmPhaseDiag}
Phase diagram of the Fermi-Hubbard model at half filling for a three-dimensional cubic lattice as a function of the particle-particle interaction $U$ and temperature $T$ in units of the tunneling energy $t$.
In the charge sector, it contains three regimes:
(i)~the Mott insulator (MI) for strong repulsive interactions and low temperature, characterized by the pinning of particle in individual lattice sites and a vanishing compressibility;
(ii)~the correlated Fermi liquid (CFL) for weak repulsive or attractive interactions, which is metallic and characterized by density fluctuations;
(iii)~the pairing regime for attractive interactions and low temperature, characterized by the formation of stable pairs of fermions with opposite spins.
These regimes are separated by smooth crossovers (dashed blue and orange lines).
In the spin sector, the competition of tunneling, interactions, and Pauli principle drives a second-order phase transition. For repulsive interactions, the transition (solid red line) separates the disordered paramagnetic (PM) phase from the anti-ferromagnetic (AFM) phase, characterized by the onset of N\'eel staggered magnetization.
For attractive interactions, the transition (solid blue line) is a normal fluid (NF) to superfluid (SF) transition, characterized by the crossover from Bose-Einstein condensate (BEC) to Bardeen-Cooper-Schrieffer (BCS) superfluidity.
}
\end{center}
\end{figure}

The phase diagram of the homogeneous Fermi-Hubbard model~(\ref{FH-Hamiltonian}) at half-filling for a simple three-dimensional cubic lattice is depicted on Fig.~\ref{fig:FHmPhaseDiag} as a function of interactions (attractive and repulsive) and temperature. The discussion is significantly simplified by separating, in a first step, the charge and spin sectors. They correspond to the spin degrees of freedom on the one hand and the motional degrees of freedom on the other hand\footnote{While in ultracold atoms, the carriers are charge-less atoms, we keep the widely used term of charges inherited from condensed-matter where the carriers are the electrically charged electrons.}.

\subsubsection{Strong repulsive interactions: Charge sector}
Let us start with the charge sector and repulsive interactions, $U>0$. At zero temperature, the repulsive interactions always dominate over the tunneling and favor the Mott insulator (MI) phase, irrespective of the value of $U/\tunnel$. This results in the pinning of a single fermion in each lattice site with an arbitrary spin, see "MI, PM" on the rightmost part of the diagram. The MI is an insulating phase, characterized by a finite gap $\Delta_\textrm{c}$, the suppression of density fluctuations, and a vanishing compressibility of the system.

For strong repulsive interactions (Mott regime), $U \gg \tunnel$, the tunneling is negligible and the fermions are strongly pinned in their lattice site. The low-energy charge excitations are particle-hole pairs. They are formed by destroying a fermion in a site, say of spin $\uparrow$, and re-creating it in another site occupied by a spin $\downarrow$ fermion. This creates a hole in the departure site and double occupation in the arrival site. The energy cost is the interaction energy and corresponds to the charge gap, $\Delta_\textrm{c} \simeq U$.

A finite temperature generates particle-hole excitations and favor density fluctuations. However, for a low temperature, $k_{\mathrm{B}}T\ll \Delta_\textrm{c}$, the MI is protected by the gap and the population of these excitations is exponentially small. For a higher temperature, $k_{\mathrm{B}}T \gtrsim \Delta_\textrm{c}$, particle-hole pairs proliferate and generate significant density fluctuations accompanied by the motion of the charges. The fermions then form a correlated Fermi liquid, see "CFL" on the top right of the phase diagram. The latter corresponds to a metallic state, characterized by a finite compressibility and charge transport.

Hence, in the charge sector, the system presents a single phase from the viewpoint of symmetry breaking but two distinct regimes with drastically different transport properties.
When the temperature increases for fixed repulsive interactions, the system crosses over from a Mott insulator to a metallic, correlated Fermi liquid at the characteristic Mott temperature
\begin{equation}\label{eq:MottT}
T_\textrm{M} \sim \Delta_\textrm{c}/\kB,
\end{equation}
see dashed orange line on Fig.~\ref{fig:FHmPhaseDiag}.
The intermediate state is a bad metal or poor insulator.
The same qualitative picture persists for weak interactions (Slater regime), $U \ll t$, down to arbitrary weak values. In this case, however, the pinning of the fermions results from the competition of interactions and tunneling and the charge gap is strongly suppressed. In contrast to the Mott regime, the charge and spin degrees of freedom are then strongly correlated and should be treated together, as first pointed out Slater~\cite{slater1951}, see Sec.~\ref{sec:FHmWI}.

\subsubsection{Strong repulsive interactions: Spin sector}\label{sec:SpinSector}

Besides the charge degrees of freedom, the Fermi-Hubbard model includes as well spin degrees of freedom. In contrast to the charges, the spin sector supports a second-order phase transition. The latter is well understood in the Mott regime, as we discuss now.

\paragraph*{Effective spin Hamiltonian~--}
For strong repulsive interactions, $U \gg \tunnel, \kB T$, the charge sector predicts a Mott insulator, as discussed above. For a vanishing tunneling, it corresponds to a Fock state with one fermion per site and an arbitrary spin $\uparrow$ or $\downarrow$. The ground level is thus an exponentially large manifold corresponding to all the possible arrangements of the $N/2$ spin $\uparrow$ and $N/2$ spin $\downarrow$ among the $N$ lattice sites.
A finite tunneling then couples these states.
Since the latter differ only by the local arrangement of the spins and the total spin is conserved, the coupling between two of these states can be seen as generated by correlated opposite spin flips in different sites. They are described by terms of the type $\hat{S}_j^+\cdot\hat{S}_\ell^-$, where $\hat{S}_j^\pm$ are, respectively, the spin raising and lowering operators at site $j$,
defined by
$\hat{S}_j^+\vert\sigma\rangle_j = \delta_{\sigma,\downarrow}(\hbar/2)\vert\uparrow\rangle_j$
and $\hat{S}_j^-\vert\sigma\rangle_j = \delta_{\sigma,\uparrow}(\hbar/2)\vert\downarrow\rangle_j$.
Physically, such terms arise from superexchange processes where two spins with opposite signs counterflow via an intermediate virtual state with one site with a hole and one site with two spins of opposite signs.
For finite but small tunneling, $\tunnel \ll U$, they may be found from second-order perturbation theory on the tunneling term. Since the Fermi-Hubbard model contains only hopping between nearest-neighbor lattice sites, it yields an effective term of the form $\hat{H}_{\textrm{I}} = \frac{\SupExch}{2} \sum_{\langle j, \ell\rangle} \big(\hat{S}_j^+\cdot\hat{S}_\ell^- + \hat{S}_j^-\cdot\hat{S}_\ell^+\big)=\SupExch\sum_{\langle j, \ell\rangle} \big(\hat{S}_j^x\cdot\hat{S}_\ell^x + \hat{S}_j^y\cdot\hat{S}_\ell^y\big)$, where
$\hat{S}_j^x=\big(\hat{S}_j^++\hat{S}_j^-\big)/2$
and $\hat{S}_j^y=\big(\hat{S}_j^+-\hat{S}_j^-\big)/2i$
are the spin operators at site $j$ in the direction $x$ and $y$.
These processes are complemented by spin-conserving terms, described by the Hamiltonian
$\hat{H}_{\textrm{II}} = \SupExch'\sum_{\langle j, \ell\rangle} \big[\hat{S}_j^z\cdot\hat{S}_\ell^z - (\hbar/2)^2\big]$.
It corresponds to processes where the system returns to the initial spin configuration via the intermediate virtual state and amounts in perturbation theory from a tunneling-induced energy shift. Notice that it vanishes when the spins in two adjacent sites are equal since the Pauli principle excludes the intermediate state, which would contain two equal spins.
Hence, up to an irrelevant constant term, the Hamiltonian reduces, in the spin sector, to the effective spin Hamiltonian
\begin{equation}\label{HeisenbergHamiltonian}
\hat{H}_\textrm{\tiny H}=J\sum_{\langle j,\ell\rangle} \mathbf{\hat{S}}_j \cdot \mathbf{\hat{S}}_\ell,
\end{equation}
corresponding to the well known Heisenberg model.
Note that rotation symmetry imposes that the superexchange amplitudes are equal in all directions, namely $J'=J$.
The complete perturbation theory treatment yields the value
\begin{equation}\label{HeisenbergJ}
J = 4t^2/U.
\end{equation}
In terms of the initial Fermi particle operators, its components along the directions $\alpha=x,y,z$ are given by
\begin{equation}\label{spins}
\hat{S}_j^{\alpha}=\frac{\hbar}{2}\begin{pmatrix}\hat{c}^{\dagger}_{j \uparrow}  &\hat{c}^{\dagger}_{j\downarrow}\end{pmatrix} \sigma_{\alpha}\begin{pmatrix} \hat{c}_{j,\uparrow}\\ \hat{c}_{j,\downarrow}\end{pmatrix},
\end{equation}
with $\hat{\mathbf{S}}=\big(\hat{S}^x,\hat{S}^y,\hat{S}^z\big)$ the vector spin operator
and $\sigma_{\alpha}$ the Pauli matrices.
In particular, the spin operator along the $z$ axis is proportional to the spin imbalance, $\hat{S}^z_j=\hbar\big(\hat{n}_{j,\uparrow}-\hat{n}_{j,\downarrow}\big)/2$, which is easily accessible experimentally (see below).

\paragraph*{Anti-ferromagnetic phase transition~--}
In the three-dimensional cubic lattice, the Heisenberg model that describes the spin sector of the Fermi-Hubbard model hosts a classical phase transition driven by thermal spin-flip fluctuations~\cite{auerbach1994}. The critical (N\'eel) temperature,
\begin{equation}\label{eq:NeelTMott}
T_\textrm{N} \sim \frac{J}{\kB} = \frac{4\tunnel^2}{\kB U},
\qquad \textrm{for} \quad U \gg \tunnel,
\end{equation}
is provided by the sole energy scale of the Heisenberg model, namely $\SupExch = 4\tunnel^2/U$, see Eqs.~(\ref{HeisenbergHamiltonian}) and (\ref{HeisenbergJ}).

For high temperature, $T \gg T_\textrm{N}$, the thermal fluctuations dominate over the exchange term. The spins attached to the lattice sites are randomly distributed and uncorrelated. The system is then in a disordered paramagnetic phase, characterized by the absence of spin ordering, see "MI, PM" on the rightmost part of the diagram.
The spin correlation function,
\begin{equation}\label{eq:SpinCorrFunct}
C_{\textrm{s}}^\alpha \big({\vec{r}}\big) = \frac{\big\langle \hat{S}^\alpha_{\vec{r}} \hat{S}^\alpha_{\vec{0}} \big\rangle - \big\langle \hat{S}^\alpha_{\vec{r}} \big\rangle \big\langle \hat{S}^\alpha_{\vec{0}} \big\rangle}{S^2},
\end{equation}
where $\vec{r}$ is a lattice site position and  $\vec{r}'=\vec{0}$ is a reference point,
decays exponentially with a spin correlation length $\xi_{\textrm{s}}$ that decreases when the temperature increases.
For low temperature, $T \ll T_\textrm{N}$, the exchange term dominates and favors the ordering of the lattice spins in a checkerboard arrangement where the spins in nearest-neighboring sites are opposed. The system is then in the anti-ferromagnetic (AFM) phase, see "MI, AFM" on the right, bottom part of the diagram. In first approximation, choosing the magnetization axis along $z$, the quantum state can be assimilated to the classical N\'eel state, $\vert\Psi\rangle_{\textrm{s}} \sim \prod_j \vert (-1)^j \rangle_j$, where $(-1)^j$ is alternatively $+1$ and $-1$ in the various lattice sites.
Since there is no privileged direction nor sign of the spin component in a particular lattice site, the choice of the magnetization axis results from spontaneous symmetry breaking. It thus fluctuates strongly from shot-to-shot realizations.
Moreover, since the spin operators in different directions do not commute, the N\'eel state cannot be an exact eigenstate of the Heisenberg Hamiltonian~(\ref{HeisenbergHamiltonian}), as can be checked easily.
Therefore, at zero temperature, the N\'eel state is dressed by quantum fluctuations, which, however, remain weak in the three-dimensional cubic lattice.
At finite temperature, they are augmented by thermal fluctuations.
Below the N\'eel temperature, they are not sufficient to destroy the anti-ferromagnetic order. The correlation function is characterized by a finite long-distance value, which decreases when the temperature increases, down to zero at the critical point.

Note that the model being short ranged with continuous symmetry, it follows from the Mermin-Wagner-Hohenberg theorem that the anti-ferromagnetic phase with long-range magnetic order is allowed in three dimensions at finite temperature and in two dimensions only at zero temperature. Therefore, the anti-ferromagnetic phase transition at the finite N\'eel temperature exists only in three dimensions. In two dimensions, the N\'eel temperature, however, provides the typical scale where the anti-ferromagnetic correlation length increases significantly.
In one dimension, only quasi-long-range anti-ferromagnetic order is possible even at zero temperature, characterized by the algebraic decay of the spin-spin correlation function.

Since the two paramagnetic and anti-ferromagnetic phases are characterized by different symmetries, one expects a second-order phase transition, in the three-dimensional case. The magnetization, $\vec{M}=\frac{2}{\hbar N}\left\langle \sum_j \hat{\vec{S_j}} \right\rangle$, where $\langle . \rangle$ denotes the quantum average over a single statistical configuration, vanishes in both phases and cannot provide an order parameter.
In contrast, the staggered magnetization,
\begin{equation}\label{eq:StaggMagn}
\vec{\mathcal{M}}=\frac{2}{\hbar N} \left\langle\sum_j (-1)^j \hat{\vec{S_j}} \right\rangle,
\end{equation}
which changes the spin sign every two sites, provides the relevant order parameter.
It vanishes in the paramagnetic phase and acquires a finite, non zero, value in the anti-ferromagnetic phase.

To conclude this paragraph, note that in the Mott regime, $U \gg \tunnel$, the N\'eel temperature~[Eq.~(\ref{eq:NeelTMott})] is much smaller than the Mott temperature~[Eq.~(\ref{eq:MottT})],
\begin{equation}\label{eq:NeelVSMottTMott}
\frac{T_\textrm{N}}{T_\textrm{M}} \sim \left(\frac{2\tunnel}{U}\right)^2 \ll 1,
\qquad \textrm{for} \quad U \gg \tunnel.
\end{equation}
Therefore, when the temperature decreases from a very high value, the system forms a Mott insulator much before the transition to the AFM phase occurs. This clear temperature-scale separation validates the separation of the charge and spin sectors, as used so far for strong interactions.

\subsubsection{Weak repulsive interactions}\label{sec:FHmWI}
The situation is more intricate in the weakly interacting (Slater) regime, $U \ll \tunnel$, where the charge and spin sectors cannot be separated any longer~\cite{slater1951}. This may inferred be from Eq.~(\ref{eq:NeelVSMottTMott}), which indicates $T_\textrm{N} \sim T_\textrm{M}$ at the crossover between the Mott and Slater regimes, $U \simeq \tunnel$. Since the charges carry a spin, the anti-ferromagnetic order cannot establish itself without freezing the charges and it is expected that the N\'eel temperature saturates to the Mott temperature,
\begin{equation}\label{eq:NeelVSMottTSlater}
{T_\textrm{N}} \simeq {T_\textrm{M}},
\qquad \textrm{for} \quad U \ll t.
\end{equation}
In other words the freezing of the charge degrees of freedom and the anti-ferromagnetic order occur simultaneously. In the Slater regime, the phase transition results from the coupled dynamics of spins and charges.
It can be described by a Hartree-Fock self-consistent decoupling and spin-charge density wave theory~\cite{slater1951,micnas1990}. This yields the critical temperature
\begin{equation}\label{eq:NeelTSlater}
T_\textrm{N} \sim \frac{6\tunnel}{\kB} \exp \left(-{7\tunnel}/{U}\right),
\qquad \textrm{for} \quad U \ll \tunnel.
\end{equation}
A similar estimate may be alternatively found from the Stoner criterion with random-phase approximation~\cite{nagaosa1999}.
It indicates that the critical temperature vanishes exponentially in the weakly interacting regime.
This is consistent with the absence of a phase transition in the non-interacting limit.
For free fermions, $U=0$, with partial filling of the ground-state Bloch band (in particular at half filling), one finds a band conductor.
The predicted behavior of the critical temperature strongly contrasts with that found in the Mott regime, where the critical temperature decreases with the interaction strength, see Eq.~(\ref{eq:NeelTMott}).
It yields a characteristic non-monotonic critical line marked by a maximum at the crossover from the Slater regime to the Mott regime, see solid red line on the phase diagram, Fig.~\ref{fig:FHmPhaseDiag}.
The latter was confirmed by numerical quantum Monte Carlo and hybrid molecular dynamics simulations~\cite{hirsch1987,scalettar1989}.

\subsubsection{Attractive interactions}

The phase diagram of the attractive case, $U<0$, is also depicted in Fig.~\ref{fig:FHmPhaseDiag}.
Similarly to the repulsive case, it contains only two phases separated by a second-order phase transition.
Here, however, the anti-ferromagnetic and paramagnetic phases are replaced by the $s$-wave superfluid (SF) and normal fluid (NF) phases. The phase diagram is remarkably symmetric with respect to the ideal Fermi gas axis, $U=0$. This is not accidental. For a bipartite lattice, this results from the partial particle-hole symmetry~\cite{shiba1972,micnas1990,ho2009a} of the model. This is established by applying the canonical transformation
\begin{equation}\label{eq:PartialParticuleHoleSymmPart}
\tilde{c}_{j,\uparrow} = \hat{c}_{j,\uparrow}
\qquad \textrm{and} \qquad
\tilde{c}_{j,\downarrow} = (-1)^j \hat{c}_{j,\downarrow}^\dagger,
\end{equation}
where $(-1)^j$ alternatively takes the values $+1$ and $-1$ in adjacent lattice sites.
In particular, the number operators read as
\begin{equation}\label{eq:PartialParticuleHoleSymmNumb}
\tilde{n}_{j,\uparrow} = \hat{n}_{j,\uparrow}
\qquad \textrm{and} \qquad
\tilde{n}_{j,\downarrow} = 1- \hat{n}_{j,\downarrow}.
\end{equation}
The transformation hence consists in transforming the spin $\downarrow$ fermions into the corresponding holes and changing the sign every two sites,
while conserving the spin $\uparrow$ fermions.
Inserting the new particle-hole operators~(\ref{eq:PartialParticuleHoleSymmPart})-(\ref{eq:PartialParticuleHoleSymmNumb}) into Eq.~(\ref{FH-Hamiltonian}), one finds $\hat{H}(+U)=\hat{H}(-U)+\textrm{cst}$. Moreover, the half-filling property is conserved, $N_{\uparrow}=N_{\downarrow}=\tilde{N}_{\uparrow}=\tilde{N}_{\downarrow}$.
The phases and regimes of the attractive Fermi-Hubbard model for a bipartite lattice at half filling are thus readily obtained from those of its repulsive counterpart.

The Mott insulator ("MI", for $U>0$) is transformed into a gas of anti-parallel spin pairs (see "pairs, NF" on the leftmost part of the diagram). The correlated Fermi liquid ("CFL" on the upper part of the diagram), which is characterized by a random distribution of spin $\uparrow$ and spin $\downarrow$ fermions among the lattice sites, is left unchanged, although the probability of pairing is enhanced in the attractive regime compared to the repulsive regime. The two regimes are connected by a smooth crossover at the characteristic temperature $T_\textrm{p} \simeq \Delta_\textrm{p} / \kB $, where $\Delta_\textrm{p}=\vert\Delta_\textrm{c}\vert$ is the pairing gap, see dashed blue line. For large, repulsive interactions, $\vert U \vert \gg \tunnel$, it is approximately the interaction energy, $\Delta_\textrm{p}\simeq \vert U \vert$.
For intermediate temperatures, so-called pseudo-gap regime, the pairs, which can be seen as randomly distributed, are incoherent and form a normal fluid (NF).
For low temperatures, long-range order establishes.
The anti-ferromagnet ("AFM", for $U>0$) is replaced by superfluid long-range order, see "SF" on the bottom, left part of the diagram.
The superfluid transition occurs at the critical temperature $T_\textrm{SF}=T_\textrm{N}$.
Degenerate with the superfluid phase, a charge-density wave phase commensurate with the lattice exists at half-filling.
For strong, attractive interactions, $\vert U \vert \gg \tunnel$, the pairs are strongly bound, and the fermions form a molecular Bose-Einstein condensate ("BEC" on the bottom, leftmost part of the diagram). In the BEC regime, superfluidity requires long-range phase coherence and the critical temperature scales as $T_\textrm{SF} \simeq 4 \tunnel^2 / \kB \vert U \vert$, which is nothing but the effective tunneling of an atom pair.
When the strength of the attractive interactions decreases, the Fermi gas crosses over to the Bardeen-Copper-Schrieffer superfluidity regime ("BCS" on the bottom, central part of the diagram) and the critical temperature scales exponentially with $\vert U \vert / \tunnel$.

\section{Quantum simulation of the Hubbard model: From ideal to strongly interacting fermions}\label{sec:FirstSimulations}
The quantum simulation of the Hubbard model using ultracold Fermi gases in optical lattices started in the early 2000's in a series of experiments. In this section, we show how standard observation tools available in ultracold atoms permit direct observation of some basic properties of Fermi lattice gases in early experiments.
We discuss signatures of Fermi statistics, for instance the appearance of a Fermi sea and the anti-bunching effect (Sec.~\ref{sec:NonInteractingFermi}), as well as the emergence of long-range phase coherence and superfluidity in interacting Fermi gases (Sec.~\ref{sec:Superfluidity}).
More advanced tools, which paved the way to further characterization of the Fermi-Hubbard model, are discussed in the next sections.

\subsection{Signatures of Fermi statistics and quantum degeneracy in ideal Fermi gases in optical lattices}
\label{sec:NonInteractingFermi}
A pioneering experiment explored the onset of the Fermi degeneracy in an ideal, single-component ultracold gas loaded in a one-dimensional optical lattice~\cite{modugno2003}. A Fermi gas of $^{40}$K atoms was prepared by sympathetic cooling with bosonic $^{87}$Rb atoms, which allowed for a direct comparison of the dynamical behaviors of the Bose and Fermi gases, in particular the absence of phase-coherence interference peaks for the Fermi gas.
In this experiment, however, the Fermi energy was much larger than the band width and a large number of Bloch bands was populated, far beyond the single-band Hubbard approach.

\paragraph*{Observation of the Fermi surface~--}
The single-band Fermi-Hubbard Hamiltonian has been realized at Zurich~\cite{koehl2005}.
The experiment is performed mostly as described in Sec.~\ref{sec:FHmUA} with an optical lattice in a simple 3D cubic  configuration. The particles are the fermionic isotope of potassium $^{40}$K in the two hyperfine states $\vert\uparrow\rangle=\vert F=9/2,\mF=-9/2\rangle$ and $\vert\downarrow\rangle=\vert F=9/2, \mF=-7/2\rangle$.
The fermions, optical pumped into the $\vert F=9/2,\mF=-9/2\rangle$ state are cooled down by sympathetic cooling with a Bose gas of $^{87}$Rb atoms. The latter is subsequently ejected exploiting the radiative pressure of a resonant laser beam and a radio-frequency (rf) pulse is used to transfer half of the $^{40}$K atoms from the $\vert F=9/2,\mF=-9/2\rangle$ state to the $\vert F=9/2,\mF=-7/2\rangle$ state, hence forming a balanced mixture of the two pseudo-spin components. After the cooling sequence, the temperature is a fraction of the Fermi temperature, $T \simeq 0.2-0.3\,\TF$ ($\TF=\EF/\kB$), and a Feshbach resonance is used to cancel the interactions between the two Fermi components.
In the presence of a harmonic trap and in the atomic limit ($\tunnel =0$), the Fermi energy is
\begin{equation}\label{eq:Fermi energy}
\EF = \frac{m\overline{\omega}^2 a^2}{2} \left(\frac{N_\sigma}{4\pi/3}\right)^{2/3},
\end{equation}
where $\overline{\omega}=(\omega_x \omega_y \omega_z)^{1/3}$ is the geometric average of the angular frequencies of the trap in the three spatial directions, $a$ is the lattice spacing, and $N_\sigma$ is the total number of fermions in the component $\sigma\in\{\uparrow,\downarrow\}$.
The filling of the lattice Brillouin zone is controlled by the dimensionless parameter $\EF/12\tunnel$, so-called characteristic filling~\cite{rigol2004}, where $12\tunnel$ is the band width in the simple cubic tight-binding model. In the experiment, it can be tuned by controlling the total atom number $N_\sigma$, the trap frequency $\overline{\omega}$, and the tunneling $\tunnel$ via the lattice amplitude, see Eq.~(\ref{Tunneling}).

The Fermi sea is imaged using a band-mapping technique. After preparing the Fermi gas at equilibrium, the optical lattice and the harmonic trap are ramped down at a rate which is slow compared to the band gap, but fast compared to the tunneling rate. The unfolded Brillouin zones convert continuously from the Bloch band structure, characteristic of the particle spectrum in the lattice, to the free-particle spectrum, with conserved quasi-momentum.
The gas then expands in free space and can be imaged as in standard time-of-flight techniques (see below).
Figure~\ref{fig:BrillouinMap} shows the 2D quasi-momentum distributions, integrated along the imaging axis ($z$), for an increasing filling factor, panels~(a) to (e).
For a weak optical lattice and low filling, the lattice potential is nearly irrelevant and the quasi-momentum distribution is spherically symmetric as in free space, although with an effective mass induced by the lattice potential.
For a strong lattice and/or high filling, the first Brillouin zone, characterized by the initial quasi-momenta $q_j \in [-\pi/a, +\pi/a]$ in each spatial direction $j\in\{x,y,z\}$, is full, and the quasi-momentum distribution, mapped onto a real momentum distribution, shows a characteristic square shape.
This result constitutes the first direct observation of the Fermi surface of a lattice Fermi gas.
The homogeneous filling of the square Brillouin zone is characteristic of the degenerate, ideal Fermi gas forming a band insulator, i.e.\ with one fermion of each spin component in each Bloch state of the lowest band.

\begin{figure}
\begin{center}
\includegraphics[width=0.90\hsize]{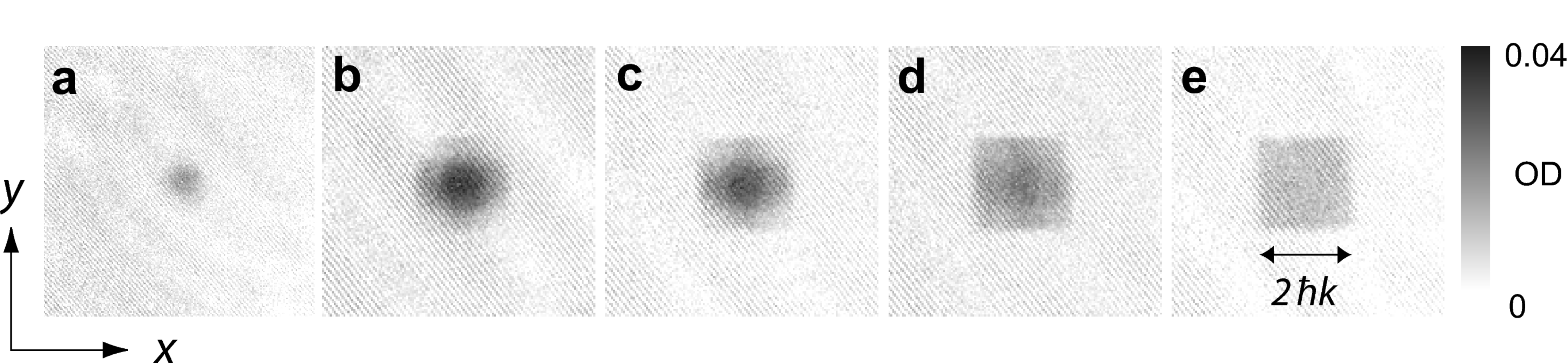}
\caption{\label{fig:BrillouinMap}
Direct observation of the Fermi sea and the first Brillouin zone of the 3D cubic tight-binding model.
The figure shows the results of band-mapping time-of-flight imaging for increasing filling rates, from panel~(a) to panel~(e).
For low filling (a), the lattice is nearly irrelevant and the Fermi surface is spherical as for free fermions with an increased effective mass.
For large filling (e), the Brillouin zone is completely full and shows a characteristic cubic shape.
Figure extracted from Ref.~\cite{koehl2005}.
}
\end{center}
\end{figure}

\begin{figure}
\begin{center}
\includegraphics[width=0.70\hsize]{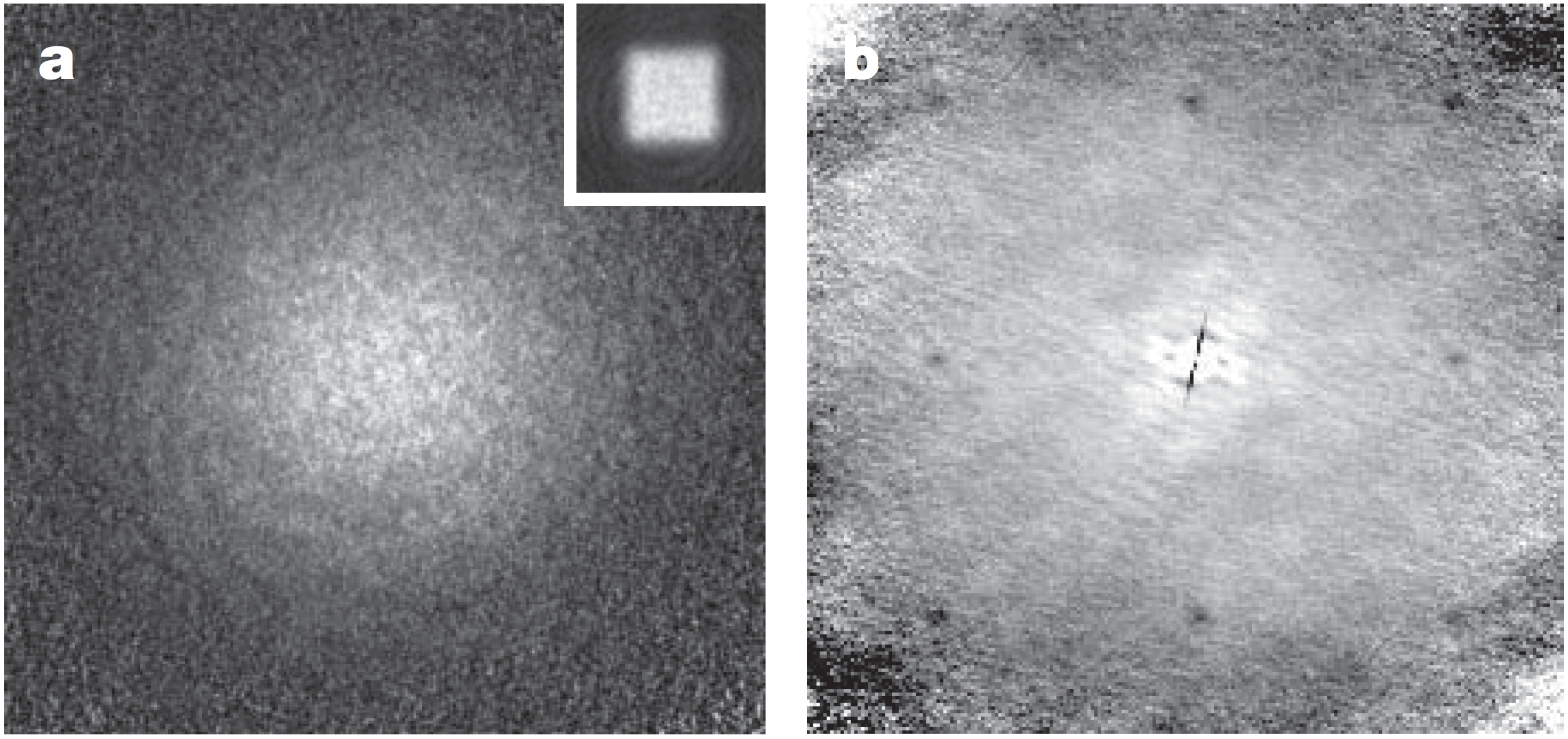}
\caption{\label{fig:NoiseSpectro}
Observation of the anti-bunching of a single-component Fermi gas in a three-dimensional optical lattice.
(a)~Time-of-flight image obtained after switching off the optical lattice instantaneously.
The inset shows the result of the same imaging but after performing band-mapping. It shows that the first Brillouin zone is fully filled, hence realizing a single-component band insulator.
(b)~Density-density correlation function as found from a series of time-of-flight images as in panel~(a). It reveals a regular array of dark spots, characteristic of the fermionic anti-bunching effect in the optical lattice.
Figure extracted and adapted from Ref.~\cite{rom2006}.
}
\end{center}
\end{figure}

\paragraph*{Fermi anti-bunching~--}
Further experiments reported other characteristic properties of Fermi lattices gases. For instance Fermi anti-bunching, characteristic of the Fermi anti-symmetrization, has been observed using noise spectroscopy analysis~\cite{rom2006}. As in the previous experiment, a single-component Fermi gas of $^{40}$K atoms is prepared in the lattice and a harmonic potential at low temperature, $T \simeq 0.2\TF$. The free fermions are distributed according to the Fermi-Dirac distribution in the Bloch states. Here only the lowest band is populated. This is revealed by the characteristic square image found from band-mapping imaging, see inset of Fig.~\ref{fig:NoiseSpectro}(a).
Due to the Pauli exclusion rule, each Bloch state is populated by at most one fermion.
The Bloch state of quasi-momentum $\hbar \vec{q}$ is a superposition of plane waves of momenta $\hbar \big[\vec{q}+\sum_j (2\pi/a)n_j \vec{\textrm{e}_j}\big]$ with $n_j\in\mathbb{Z}$ and $j \in\{x,y,z\}$ spans the spatial directions. Since $\vec{q}$ is by definition in the first Brillouin zone, $q_j\in [-\pi/a,+\pi/a]$, the components of two different Bloch states are all different. It follows that if a fermion is detected in the real-momentum state $\hbar\vec{k}$, no fermion can be detected in any momentum state $\hbar \big[\vec{k}+\sum_j (2\pi/a)n_j \vec{\textrm{e}_j}\big]$ with $n_j\in\mathbb{Z}$. To observe this anti-bunching effect, the experimentalists have performed time-of-flight imaging. The latter is one of the most standard measurement tools in ultracold atoms. It consists in cutting off abruptly all confining fields.
When the interactions are negligible (as in a polarized Fermi gas), the
atoms expand in free space and the momentum distribution is conserved. The ballistic dimension is quasi-classic and each atom travels a distance proportional to its initial velocity. Waiting a time larger than the cloud size divided by the characteristic velocity, the initial position distribution is irrelevant and the distance is directly proportional to the initial velocity.
It directly maps the initial momentum distribution onto the expanded position distribution. The latter can be imaged by absorption images, which integrate the atomic distribution along the laser axis.

In contrast to the previous experiment, the time-of-flight sequence is performed without band-mapping, i.e.\ by switching off the lattice and trap instantaneously. Each image shows an approximate Gaussian-like shape, characteristic of the expanded Wannier function, with no hole, see Fig.~\ref{fig:NoiseSpectro}(a).
From the images of a large number of independent realizations of the same experiment, one then computes the density-density correlation function
\begin{equation}\label{eq:FermiCorrelations}
C(\vec{\Delta r}) = \frac{\int d\vec{r}\, \big\langle \hat{n}(\vec{r}-\vec{\Delta r}/2) \hat{n}(\vec{r}+\vec{\Delta r}/2)\big\rangle}{\int d\vec{r}\, \big\langle \hat{n}(\vec{r}-\vec{\Delta r}/2)\big\rangle \big\langle \hat{n}(\vec{r}+\vec{\Delta r}/2)\big\rangle},
\end{equation}
at a variable separation $\vec{\Delta r}$ after time-of-flight.
For an ideal gas classical gas, one finds $C(\vec{\Delta r})=1$ since the particles are independent. In contrast, a value $C(\vec{\Delta r})<1$ is characteristic of Fermi anti-bunching, while a value $C(\vec{\Delta r})>1$ is characteristic of Bose bunching.
The result shows clear dark spots that can be mapped onto the quantities $\sum_j (2\pi/a)n_j \vec{\textrm{e}_j}$ in the initial momentum distribution, see Fig.~\ref{fig:NoiseSpectro}(b). This is the clear signature of  the anti-bunching effect, characteristic of the anti-symmetrization of fermionic states.

\subsection{Phase coherence of an interacting Fermi gases in an optical lattice}\label{sec:Superfluidity}
Signatures of coherence for a strongly interacting Fermi gases in an optical lattice were reported in Ref.~\cite{chin2006}. The experiment shares most of its features with the previous ones, but interactions are set on.
It exploits two hyperfine states of a fermionic isotope, here $^6$Li atoms. A Feshbach-resonance technique is used to adjust the scattering length to a large value and create tightly bound Fermi pairs with a size smaller than the lattice spacing. Both sides of the BEC-BCS crossover have been explored. A condensate of Fermi pairs is first produced in a harmonic trap and then loaded in the three-dimensional optical lattice. To probe the lattice gas, the Feshbach magnetic field is then tuned far away from the resonance, which drastically lowers the interaction strength.
A time-of-flight image is then realized by switching off all lattice laser beams almost instantaneously.

\begin{figure}
\begin{center}
\includegraphics[width=1.0\hsize]{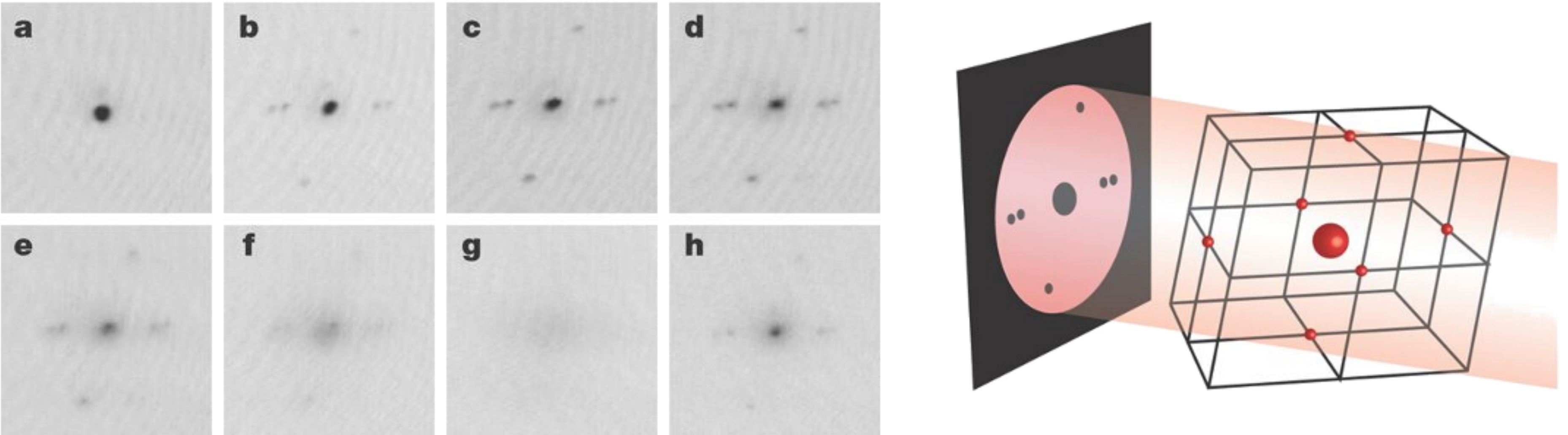}
\caption{\label{fig:FermiSF}
Observation of phase coherence in a strongly interacting Fermi gas in an optical lattice.
Panels (a)-(g) show time-of-flight images of the Fermi gas released from the lattice for increasing lattice amplitudes:
(a)~$V_0=0\Er$,
(b)~$V_0=2.5\Er$,
(c)~$V_0=4\Er$,
(d)~$V_0=5\Er$,
(e)~$V_0=6\Er$,
(f)~$V_0=7\Er$,
and (g)~$V_0=9\Er$.
Panel~(h) shows the result for $V_0=2.5\Er$ as in panel~(b) but taken after first ramping up the lattice to $V_0=10\Er$ before ramping it down to $V_0=2.5\Er$.
Figure extracted and adapted from Ref.~\cite{chin2006}.
}
\end{center}
\end{figure}

In the absence of the lattice, one observes a spherical, structureless pattern, characteristic of the momentum distribution in free space, see Fig.~\ref{fig:FermiSF}(a). For intermediate values of the lattice amplitude $V_0$, a distinct interference pattern appears, see Fig.~\ref{fig:FermiSF}(b)-(e). This interference pattern is the signature of long-range phase coherence. On the images, one observes six interference peaks. In the experiment, the imaging beam was inclined with respect to the lattice axes, see right panel of Fig.~\ref{fig:FermiSF}. More precisely, it was nearly $45^\circ$ off axis in the horizontal plane and the integrated absorption image produces two pairs of peaks close to each other along the horizontal line of the image. As expected, taking into account the geometry of the imaging system, the observed absorption peaks could be mapped onto the first-order Bragg peaks of the optical lattice at $q = \pm 2\hbar \kL$ in each spatial direction.

For larger lattice amplitudes, the interference pattern is washed out, see Fig.~\ref{fig:FermiSF}(f)-(g).
It is the signature of an insulator regime.
By ramping up the lattice to a large value, waiting a transient time, and then ramping down the lattice to a smaller value, the interference pattern is recovered, see Fig.~\ref{fig:FermiSF}(h).
The reversibility of the process permits to rule out decoherence or temperature effects as responsible for the disappearance of the interference pattern in the strong lattice. For repulsive interactions, it may be attributed to the onset of a molecular Mott insulator. For attractive interactions, it may be related to a band insulator. However, in the experiment, the filling was close to unity. Moreover, the interaction energy exceeded the band gap and a multi-band model, beyond the scope of this review, was necessary to describe the physics~\cite{zhai2007,moon2007}.

The observation of an interference pattern is the signature of the onset of phase coherence in the interacting lattice Fermi gas. The latter may be attributed to superfluidity. However, no direct proof of superfluidity has been reported in a lattice Fermi gas so far. In principle, this may be realized via the observation of vortices in rotating gases or by studying the response of the gas to a moving obstacle, as done for homogeneous Bose and Fermi gases~\cite{raman1999,madison2000,zwierlein2005,desbuquois2012}.

Subsequent experiments in Zurich, Mainz, and Princeton have explored attractive Fermi-Hubbard systems in the single-band regime by employing more moderate values of the interaction strength $U$~\cite{strohmaier2007,hackermueller2010,schneider2012,mitra2018}. Although the temperatures were too high to observe superfluid behavior and long-range coherence, the latest Princeton experiments could measure precursors of the superfluid phase~\cite{mitra2018}.

\section{Observation of the fermionic Mott insulator}\label{sec:observMott}
Soon after the realization of degenerate Fermi gases in optical lattices, strong efforts were carried out to enter the strongly correlated regime.
In this section, we first review pioneering experiments, which demonstrated the onset of a Mott insulator state in bulk measurements~(Sec.~\ref{sec:MIbulk}).
We then discuss a second generation of experiments, which exploit spatially resolved, in-situ imaging and allow for the distinction of coexisting metallic and insulating states in inhomogeneous traps~(Sec.~\ref{sec:spatiallyresolved}).
We finally discuss the case of two-dimensional lattices, which are particularly suited for such measurements~(Sec.~\ref{sec:2D}).

\subsection{First observations of the fermionic Mott insulator}\label{sec:MIbulk}

The fermionic insulating Mott regime, characteristic of the charge sector of the Hubbard model, has been first attained in 2008 in two experiments at Zurich~\cite{jordens2008} and Munich~\cite{schneider2008}.
Various signatures have been demonstrated using an independent control of the hopping amplitude, the interaction strength, and the harmonic trap potential.
The setup is similar to that used in Ref.~\cite{koehl2005}, see Sec.~\ref{sec:NonInteractingFermi}.

\begin{figure}
\begin{center}
\includegraphics[width=1.0\hsize]{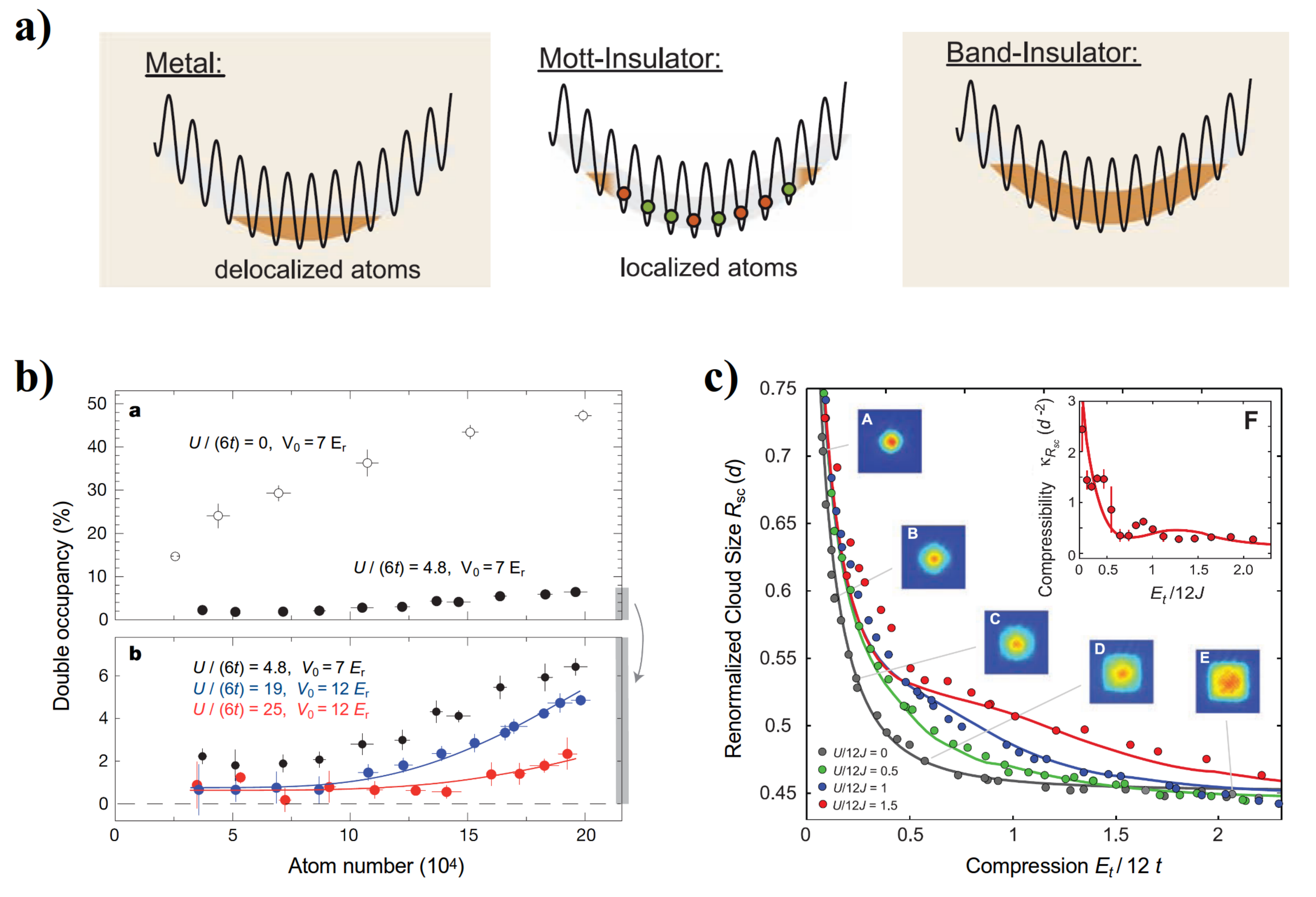}
\caption{\label{fig:obsMott}
Experimental observation of the Mott insulator in ultracold-atom realizations of the Fermi-Hubbard model, in a harmonic trap.
(a)~Schematic representation of the various regimes found from the competition of the band width $12t$, the repulsive interaction strength $U$, and the free-gas Fermi energy $\EF$:
Metal ($U \ll \EF \ll 12 t$, left panel), Mott insulator ($\EF \ll 12t \ll U$, central panel), and band insulator ($\EF \gg U, 12t$, right panel).
(b)~Demonstration of the suppression of double occupancy. The figure shows the fraction of doubly occupied lattice sites as a function of the total atom number for various interaction energies:
$U/6t=0$ (open points), $U/6t=4.8$ (filled black points), $U/6t=19$ (blue points), and $U/6t=25$ (red points).
The lines correspond to theoretical predictions for the trapped system in the atomic ($t=0$) limit, which is the strongly interacting regime, and at the temperatures realized experimentally. They are nearly inextinguishable from those obtained from high temperature series expansions or DMFT.
(c)~Demonstration of the suppression of the compressibility. The main panel shows the size of the Fermi gas as a function of the trap-induced compression, as characterized by the Fermi energy in units of the lattice band width, for various interaction strengths (black, green, blue, and red data). The insets A-E show measurements of the quasi-momentum distribution for the non-interacting gas, showing the progressive filling of the first Brillouin zone.
The inset~F shows the heuristic compressibility as a function of the compression, showing a minimum, characteristic of the Mott insulating core. The points show the experimental data and the solid lines the corresponding DMFT numerical results.
The various panels are extracted from Refs.~\cite{jordens2008,schneider2008}, and adapted.
}
\end{center}
\end{figure}

The harmonic trap plays a central role and permits to reach half filling in the trap center for certain parameters. The state of the two-component Fermi gas is determined by the competition of three characteristic energies,
namely the single-particle band width $12\tunnel$, the repulsive interaction strength $U$, and the Fermi energy $\EF$ of the free gas in the atomic limit ($t=0$), see Eq.~(\ref{eq:Fermi energy}).
For $U \ll \EF \ll 12 \tunnel$, the interactions are negligible and one finds a delocalized Fermi sea, corresponding to a metal, see left panel on Fig.~\ref{fig:obsMott}(a).
For $\EF \ll 12\tunnel \ll U$, one expects a Mott regime where each lattice site is occupied by one and only one fermion, either in the spin $\uparrow$ or spin $\downarrow$ state. The ground state is formed by first populating the lattice sites in the trap center and progressively filling peripheral sites, which have a larger on-site harmonic energy, see Eq.~(\ref{TrappingTerm}). In dimension higher than one, the last spherical shell may be partially filled and forms a metallic region, see central panel on Fig.~\ref{fig:obsMott}(a).
For $\EF \gg U, 12\tunnel$, it becomes energetically more favourable to form doubly occupied sites in the center rather than new peripheral sites. One then forms a band insulator with a spin $\uparrow$ and a spin $\downarrow$ fermion in each lattice site near the trap center, see right panel on Fig.~\ref{fig:obsMott}(a).

\paragraph*{Suppression of doubly occupied sites~--}
The experiment at Zurich revealed the characteristic suppression of density fluctuations in the Mott-insulating phase, determined by spectroscopic measurements of the fraction of doubly occupied sites of the complete system~\cite{jordens2008}.
The experiment actually exploits three states of the $F=9/2$ hyperfine level of $^{40}$K atoms. The two states $\vert\uparrow\rangle=\vert F=9/2, \mF=-9/2\rangle$ and $\vert\downarrow\rangle=\vert F=9/2, \mF=-5/2\rangle$ here encode the two spin states of the Hubbard model~(\ref{FH-Hamiltonian}), while the third state, $\vert F=9/2, \mF=-7/2\rangle$ is used as an ancilla for spectroscopic measurements.
In brief, the measurement of the fraction of doubly occupied sites uses a radio-frequency (rf) pulse that transfers the atoms from the $\vert\downarrow\rangle=\vert F=9/2, \mF=-5/2\rangle$ state to the ancilla state $\vert F=9/2, \mF=-7/2\rangle$. The rf-pulse frequency is adjusted to be resonant with the transition energy shifted by the on-site interaction strength $U$. Hence, only the fermions initially in the $\vert\downarrow\rangle=\vert F=9/2, \mF=-5/2\rangle$ state and in doubly occupied sites are transferred to the state $\vert F=9/2, \mF=-7/2\rangle$. The total number of fermions in this state are then measured after a resonant time-of-flight sequence. This directly yields the number of doubly occupied sites~\cite{jordens2008}.

Figure~\ref{fig:obsMott}(b) shows the fraction of doubly occupied sites versus the total atom number in the balanced case, $N_\uparrow=N_\downarrow$, for various values of the interaction strength.
For free fermions ($U=0$, open points), the fraction of doubly occupied sites increases rapidly with the total number of fermions.  It may be interpreted as the partial filling of the lattice sites, independently for the two spin components. This is characteristic of the metallic state, see left panel in Fig.~\ref{fig:obsMott}(a).
For significant repulsive interactions ($U \gg \tunnel$, filled points), the fraction of doubly occupied sites is strongly suppressed. It is nearly frozen for small atom numbers. This is characteristic of the Mott regime, see central panel of Fig.~\ref{fig:obsMott}(a). As expected, the suppression of double occupation is stronger when the interaction strength increases. When the total number of atoms further increases, the fraction of doubly occupied sites increases, but significantly less than for free fermions. This is compatible with the formation of a Mott-insulating core in the trap center, surrounded by a metallic shell.
For an increasing atom number, doubly occupied sites start to form in the center of the trap and are responsible for the increase of double occupancy depicted in the right panel of Fig.~\ref{fig:obsMott}(a).

Quantitative comparisons of the experimental data to various theoretical approaches were performed in a second series of experiments~\cite{jordens2010}.
A high-temperature series expansion (HTSE) of the grand partition function, where $t/\kB T$ is the small parameter, is particularly useful to describe the thermodynamic properties of the system in the temperature regime where the experiments were performed~\cite{scarola2009}. The predictions of the HTSE are in excellent agreement with DMFT simulations~\cite{deleo2011}, and with experiments performed in a broad range of parameters encompassing the metallic and Mott-insulating regimes~\cite{jordens2010}.

The formation of a Mott-insulating core was confirmed by the observation of the finite Mott gap. A periodic, time-dependent, weak ($\sim 10\%$) modulation of the lattice amplitude was exploited in the same experiment to probe the excitation spectrum of the system~\cite{jordens2008}. It couples the ground state to excited states with an energy determined by the modulation frequency $\nu$, which is ajustable in the experiment.
The response of the system is measured via the increase of the fraction of doubly occupied sites. For sufficiently large interaction strength, it shows a marked peak at $\nu \simeq U/h$ with $h$ the Planck constant, characteristic of the Mott gap in the strongly interacting regime.
Nearest-neighbor density and spin correlation functions in the linear response regime have also been determined using an extension of the same approach~\cite{greif2011}.
Site-occupation and lattice modulation measurements were subsequently performed in other lattice configurations, for instance honeycomb lattices~\cite{uehlinger2013}, a ionic version of the Fermi-Hubbard Hamiltonian~\cite{messer2015}, and with alkali-earth atoms. The latter gives access to systems with a large number of internal states ($\kappa=6$) and, correspondingly, to $SU(6)$ fermionic Mott insulators~\cite{taie2012}.

\paragraph*{Suppression of the compressibility~--}
The experimentalists at Mainz observed the suppression  the compressibility, which provides another signature of the Mott insulating state~\cite{schneider2008}. The main difference compared to the Zurich experiment is the use of a red-detuned dipole trap, independent of the (blue-detuned) optical lattice, to probe the response of the Fermi gas to a compression induced by increasing the harmonic trap frequency.
The trap size $R = \sqrt{\langle r^2\rangle}$ is extracted from direct measurement of the density profile using in-situ phase-contrast imaging of the Fermi gas in the presence of the lattice and the harmonic dipole trap. The resulting reduced size, $R_\textrm{sc} \propto R/N_\sigma^{1/3}$ is plotted versus the dimensionless Fermi energy $\EF/12\tunnel$ on Fig.~\ref{fig:obsMott}(c) for various values of the interaction strength. The value of Fermi energy, $\EF \propto \overline{\omega}$ [see Eq.~(\ref{eq:Fermi energy})], provides a measure of the compression induced by the trap.
For vanishing interactions (black points), the compression induces the progressive filling of the first Brillouin zone, as observed by band-mapping time-of-flight imaging of the quasi-momentum distribution (see inset panels A to E), similarly as discussed in Sec.~\ref{sec:NonInteractingFermi}. As a result size of the gas decreases continuously when the compression increases, and saturates when $\EF \gtrsim 12\tunnel$ as observed in the experiment.
For increasing repulsive interactions (color points, from green to red), the response of the gas is suppressed and for sufficiently large interactions, one finds a quasi-plateau, where the response almost vanishes (see red curve, $U/12\tunnel=1.5$, for $0.5 \lesssim \EF/12\tunnel \lesssim 0.7$). This is compatible with the formation of a Mott-insulating core, surrounded by a metallic spherical shell. For larger interactions, $\EF \gtrsim U, 12\tunnel$, the Fermi pressure exceeds the interactions and the size of the gas approaches that found in the absence of interactions, indicating the formation of a band insulator. These global measurements offer a signature of the Mott insulating regime and complement the local suppression of doubly occupied sites discussed above.

Data such as those of Fig.~\ref{fig:obsMott}(c) allows to derive the compressibility of the interacting Fermi gas. Here the harmonic trap frequency plays the role of the pressure and one may use the heuristic definition
\begin{equation}\label{eq:compressibility}
\kappa_{R_\textrm{sc}} = - \frac{1}{R_\textrm{sc}^3} \frac{\partial R_\textrm{sc}}{\partial (\EF/12\tunnel)}.
\end{equation}
The result, shown in the inset~F of Fig.~\ref{fig:obsMott}(c) for strong interactions, $U/12\tunnel=1.5$, displays a characteristic minimum versus compression. It confirms the formation of a Mott insulating core.

All the experimental data have been compared to large-scale, ab initio, numerical calculations using dynamical meanfield theory (DMFT)~\cite{georges1996} within the local density approximation (LDA).
Very good agreement between the experimental data and the numerical calculations was found without fitting parameter, see solid lines on all panels of Fig.~\ref{fig:obsMott}(c). This holds up to the maximum interaction strength explored in the experiment, $U /12\tunnel \simeq 1.5$, although slight deviations are observed for the largest values. This cross-validates the experiment as a quantum simulator and the accuracy of the DMFT approach for strongly correlated Fermi gases.

\subsection{Spatially resolved experiments}\label{sec:spatiallyresolved}

In all the previous experiments, global measurements over the complete system have provided evidence of the emergence of a Mott insulator core from bulk quantities, encompassing also the metallic, non half filled, shells. Further experiments have been realized exploiting instead \emph{in situ} images of the atomic cloud density profile. Using a sufficient spatial resolution, it gives selective access to the half-filled region, and hence to a pure Mott insulator.

For three-dimensional lattices, it is necessary to apply a reconstruction scheme.
Absorption imaging integrates the density profile $n(x,y,z)$ along the imaging laser, say along axis $z$.
It yields the so-called column density $\overline{n}(x,y)=\int dz\, n(x,y,z)$, which, in general, contains less information than the full density profile. Assuming cylindrical symmetry, $n(x,y,z)=n\big(x,\sqrt{y^2+z^2}\big)$, however, the three-dimensional density profile can be reconstructed from the column density using an inverse Abel transformation~\cite{bracewell1965,dribinski2002}.

Such an approach has been developed to study the standard Fermi-Hubbard model in a two-component Fermi gas of $^6$Li atoms~\cite{duarte2015}.
For sufficiently large repulsive interactions and a smooth enough harmonic trap, the reconstructed on-site density profile shows a clear plateau at the on-site density $n=1$ around the trap center and a smooth decay at a larger distance.
This local measurement permits to distinguish the Mott insulator core in the center from the metallic shell in the periphery of the trap.
It further allows to derive the local, isothermal compressibility per lattice site volume, $\kappa = n^{-2} {\partial n}/{\partial \mu}$, from the density profile.
To do so, the local density approximation (LDA) is used.
For a spherically symmetric trap of angular frequency $\omega$, it yields the local chemical potential $\mu(r)=\mu-m\omega^2 r^2 /2$, where $\mu$ is the global thermodynamic chemical potential and $r=\sqrt{x^2+y^2+z^2}$ is the radial distance. Inserting the expression for $\mu(r)$ into the definition of the compressibility, one then finds
\begin{equation}\label{eq:LocalCompressibility}
\kappa(r) = \frac{1}{m\omega^2} \times \frac{1}{n^2\, r} \frac{\partial n}{\partial r}.
\end{equation}
Remarkably, the local compressibility does not depend on the global chemical potential $\mu$. It is directly derived from the density profile. The local compressibility $\kappa(r)$ is a function of the radial distance $r$, or equivalently of the density $n(r)$.
The relation between $\kappa$ and $n$ yields the equation of state of the homogeneous gas, locally, at the density $n$.
Hence, within the LDA, the harmonic trapping gives access to the thermodynamics of the homogeneous gas as a function of the density in a single experiment. The experimental results show a finite compressibility for any on-site density $n<1$, which vanishes at unity, $n=1$. This is consistent with the suppression of compressibility in the Mott insulating state, now revealed at the local level.

A similar study has been performed on the Fermi-Hubbard model extended to a particle spin degeneracy higher than $2$~\cite{hofrichter2016}. The experiment exploits $^{173}$Yb atoms. The atomic ground state has the nuclear angular momentum $I=5/2$ and no electronic angular momentum, $J=0$.
The 6 nuclear angular-momentum states $\vert I=5/2, m_I\rangle$ with $m_I=-5/2, -5/2+1, ..., 5/2$ encode the $5/2$-spin states of the extended Hubbard model.
Due to the decoupling of the nuclear and electronic spins, the interaction is independent of the nuclear spin state and the system is $SU(\kappa)$ symmetric with $\kappa=2I+1=6$.
The $SU(3)$-symmetric case is also realized in the same experiment by optical pumping of the atoms out of the 3 unwanted spin states.
In both cases, Mott plateaus at the total on-site density $n=1$, accompanied by a vanishing compressibility, are observed for sufficiently large interaction strength. The equation of state of the homogeneous system $n(\mu,T)$ was then determined from the density profile within the LDA.

\subsection{Two-dimensional Fermi-Hubbard model}\label{sec:2D}
Two-dimensional systems are more favourable for local measurements for
each \emph{in situ} absorption image directly yields the two-dimensional density profile of the gas $n(x,y)$, without the need of any reconstruction schemes. An accurate knowledge of the external trapping potential $V(x,y)$ and of the temperature is sufficient to obtain the equation of state of the system, $n(\mu,T)$.
A series of experiments on the two-dimensional Fermi-Hubbard model has been performed using $^{40}$K atoms.
Demonstrated in Ref.~\cite{cocchi2016}, this method also allows to obtain the temperature $T$ of the system by comparing the measured equation of state at low-densities to the theoretical predictions of a numerical linked-cluster expansion~\cite{khatami2011}.

\begin{figure}
\begin{center}
\includegraphics[width=0.95\hsize]{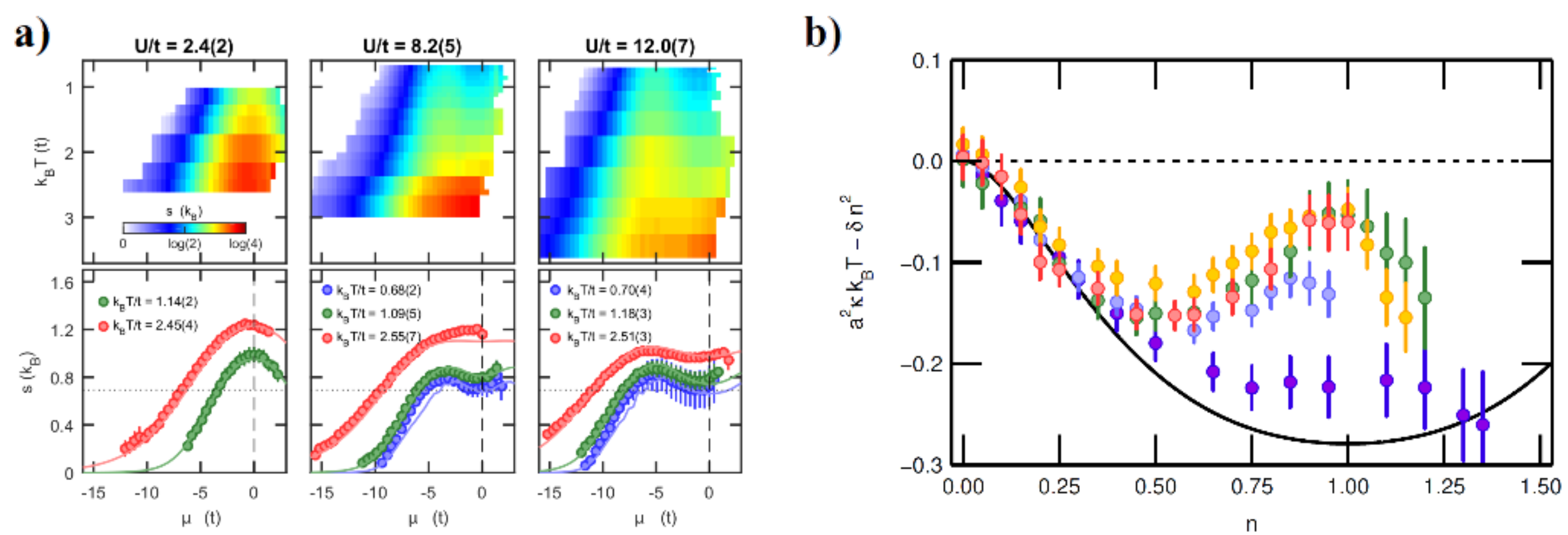}
\caption{\label{fig:entropy-compressibility}
In-situ signatures of the Mott crossover in the two-component Fermi gas in the two-dimensional Hubbard model.
(a)~Measurement of the entropy.
The various columns correspond to increasing values of the interaction strength $U/\tunnel$.
The upper row shows, in color scale, the entropy per lattice site determined by Eq.~(\ref{eq:entropy}) as a function of the chemical potential $\mu$ and the temperature $T$.
The lower row shows cuts of corresponding figures on the upper row for various fixed temperatures.
Half filling is marked by the vertical dashed line at $\mu=0$.
Figure extracted from Ref.~\cite{cocchi2017} (note that it has been adapted to match the notations of the present paper for the chemical potential, $\mu - U/2 \rightarrow \mu$).
(b)~Measurement of non-local density fluctuations.
The quantity $\sum_{\ell \neq j} G_{0,\ell}$ is plotted as a function of the on-site lattice density $n$ for various interaction strengths,
$U/\tunnel \simeq 1.6$ (purple)
$U/\tunnel \simeq 6.1$ (light blue),
$U/\tunnel \simeq 8.2$ (green,),
$U/\tunnel \simeq 10.3$ (yellow),
and $U/\tunnel \simeq 12.0$ (red).
The solid line is the prediction for the ideal, trapped Fermi gas at $\kB T/\tunnel=0.65$.
Figure extracted from Ref.~\cite{drewes2016}
}
\end{center}
\end{figure}

In subsequent experiments, in-situ imaging of density profiles taken at different temperatures have been exploited to determine various thermodynamic properties of the system~\cite{cocchi2017}.
For instance, in a hydrodynamic approach within the LDA, the local pressure $P(\vec{r})$ is found from the dynamical equilibrium condition $\vec{\nabla}P(\vec{r})+n(\vec{r}) \vec{\nabla}V_{\textrm{trap}}(\vec{r})=0$, where $V_{\textrm{trap}}(\vec{r})$ is the harmonic trapping potential. Assuming rotational symmetry and integrating this equation from the radial distance $r$, where the effective, local chemical potential is here noted $\mu$, to the region of vanishingly small density, one finds the pressure of the homogeneous systems,
\begin{equation}\label{eq:pressure}
P(\mu,T) = m\omega^2 \int_{r}^\infty dr'\, r' n(\vec{r}'),
\end{equation}
where the density profile is measured at the fixed temperature $T$.
The entropy per lattice site, $s=(a^2/\Sigma)\times S$, where $S$ is the total entropy, $\Sigma$ is the system area, and $a$ is the lattice spacing, is then found using the Gibbs-Duhem relation $SdT-\Sigma dP+Nd\mu=0$. For a fixed chemical potential, it yields
\begin{equation}\label{eq:entropy}
s = a^2 \left.\frac{\partial P}{\partial T}\right\vert_\mu.
\end{equation}
The derivation is performed numerically from the experimental data taken at different temperatures with a fixed chemical potential.
For illustration, Fig.~\ref{fig:entropy-compressibility}(a) shows the results of such measurements.
The various columns  correspond to increasing values of the interaction strength in units of the tunneling energy, $U/\tunnel$. For low interactions [left column], the entropy peaks at half filling ($n=1$, vertical dashed line). On the contrary, for larger interactions [central and right columns], the entropy shows a marked minimum at half filling, which is characteristic of the Mott charge gap.
This result can be understood intuitively. In the Mott regime, the value of the entropy per site is $s=\kB \ln 2$. This corresponds to the fact that, for a site of occupation $1$, two possible spin configurations ($\uparrow$ and $\downarrow$) exist. For lower values of the chemical potential (and thus of the occupation), there is a local maximum of $s$ because three configurations (spin $\uparrow$, spin $\downarrow$, and hole) are possible. 

Comparison between thermodynamic observables and local quantities even permits to demonstrate the existence of non-local correlations in the system~\cite{drewes2016,cocchi2017}. In situ imaging give access to the density correlations of the interacting Fermi gas,
\begin{equation}\label{eq:DesnCorr}
G_{j,\ell} = \big\langle \hat{n}_j \hat{n}_\ell \big\rangle - \big\langle \hat{n}_j \big\rangle \big\langle \hat{n}_\ell \big\rangle,
\end{equation}
with $\hat{n}_j = \hat{n}_{j,\uparrow} + \hat{n}_{j,\downarrow}$ the total atom number on the lattice site $j$.
In a conducting state, such as that obtained with a partially filled Bloch band, the atoms are delocalized and establish long-range order, corresponding to $G_{j,\ell}<0$ due to Fermi anti-bunching.
In localized states, such as the band insulator or the Mott insulator, one expects $G_{j,\ell} \simeq 0$ since all the lattice sites are frozen. Measuring directly the density correlation function~(\ref{eq:DesnCorr}) is difficult with standard imaging, where the resolution exceeds one lattice site. To overcome this issue, it is possible to exploit the thermodynamic relation between the (reduced) compressibility $\tilde{\kappa} = a^{-2}{\partial \langle N \rangle}/{\partial \mu}$ and the atom number fluctuations $\Delta N^2 = \big\langle \hat{N}^2 \big\rangle - \big\langle \hat{N} \big\rangle^2$, or, equivalently, the zero-momentum static, density structure factor, $S_n(\vec{q}=0)=\Delta N^2/\langle N \rangle$.
At the local level ($\tilde{\kappa}=\sum_j\tilde{\kappa}_j$), it reads as
\begin{equation}\label{eq:DesnCorrCompress}
\tilde{\kappa}_j = \frac{1}{a^2 \kB T} \Big( \Delta n_j^2 + \sum_{\ell \neq j} G_{j,\ell} \Big).
\end{equation}
In the experiment of Ref.~\cite{drewes2016}, the non-local correlations from the lattice site $j$, $G_{j,\ell}$ are found from the independent measurements of the reduced local compressibility $\tilde{\kappa}_j$ and of the on-site density fluctuations $\Delta n_j^2$.
The local compressibility is found from the spatial derivative of the density profile as discussed for the three-dimensional case above. The local density fluctuations are found from the relation
$\Delta n_j^2 = 2 \big\langle \hat{n}_{j\uparrow} \big\rangle - 4 \big\langle \hat{n}_{j\uparrow} \big\rangle^2 + 2 \big\langle \hat{n}_{j\uparrow} \hat{n}_{\ell\downarrow} \big\rangle$, derived by using the Fermi anti-commutation rules, and the combination of in situ images (yielding the value of $\big\langle \hat{n}_{j\uparrow} \big\rangle$) and space-resolved radio-frequency spectroscopy  (yielding the value of $\big\langle \hat{n}_{j\uparrow} \hat{n}_{\ell\downarrow} \big\rangle$). Figure~\ref{fig:entropy-compressibility}(b) shows experimental results for the non-local density fluctuations at the trap center, $\sum_{\ell \neq j} G_{0,\ell}$, as a function of the on-site atomic density for various interaction strengths.
For weak interactions (purple points), the data are close to the theoretical prediction for the ideal, trapped Fermi gas (solid black line). For increasing interactions (other color points), the non-local density fluctuations are also close to the ideal case for low density but show strong deviations when the density increases. The data at half-filling, $n=1$, approach the prediction for insulating states, $\sum_{\ell \neq j} G_{0,\ell} \simeq 0$, indicating the onset of a Mott insulator.

\begin{figure}
\begin{center}
\includegraphics[width=0.95\hsize]{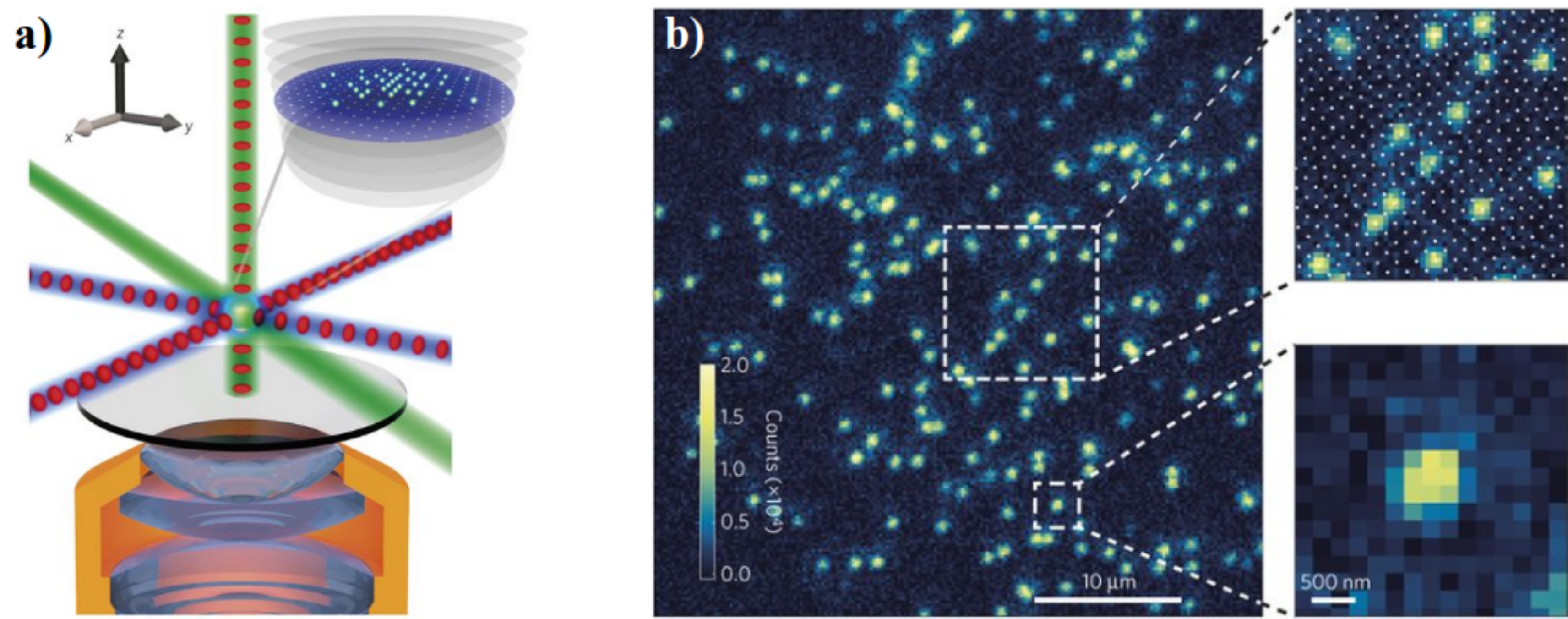}
\caption{\label{fig:microscope}
Quantum gas microscope.
(a)~Experimental scheme. An optical lattice (red dots) is realized using three retro-reflecting beams, creating standing waves in directions. The green and blue beams represent additional laser beams used for electromagnetically induced-transparency cooling in the lattice sites. Fluorescence photons are then collected in a high-resolution microscope, oriented vertically (orange and gray).
It produces single-site resolved images of the atoms in the focal plane of the microscope.
(b)~Fluorescence image of $^{40}$K atoms in an optical lattice.
Each light spot represents a single atom, with a resolution of the order of $500\mu$m (see lower inset).
The positions of the atoms can be mapped onto the periodic array of lattice sites (see white dots on the upper inset).
Figures extracted from Ref.~\cite{haller2015}.
}
\end{center}
\end{figure}

\paragraph*{Experiments under the quantum gas microscope~--}
A key step in the study of the Fermi-Hubbard model has been the development of quantum-gas microscopes. This technique is well adapted to lattice systems with a few atoms per lattice site and provides in-situ images with a resolution at the level of one lattice site.
Pioneered for bosonic atoms in 2009~\cite{bakr2009,sherson2010}, this technique has become available for fermionic atoms only in 2015~\cite{haller2015,cheuk2015,parsons2015,omran2015,edge2015}.
In brief, the technique works as follows.
After the preparation of the Fermi gas at equilibrium in the optical lattice, the amplitude of the latter is suddenly turn to a large value so that quantum tunneling is strongly suppressed. The atoms are then frozen in deep lattice wells, which may accommodate many bound states. Laser cooling techniques, such as electromagnetically induced-
transparency cooling~\cite{morigi2000} or Raman side-band cooling~\cite{hamann1998}, are applied. They not only bring the atoms down to the most bound state so as to minimize the losses but also enhance the emission of fluorescence photons. The latter is then collected with a high-resolution microscope focused on a lattice plane, see Fig.~\ref{fig:microscope}(a). With a resolution lower than the lattice spacing, it provides direct images on the atom distribution, see Fig.~\ref{fig:microscope}(b), which can be mapped on the lattice site positions.

\begin{figure}
\begin{center}
\includegraphics[width=0.95\hsize]{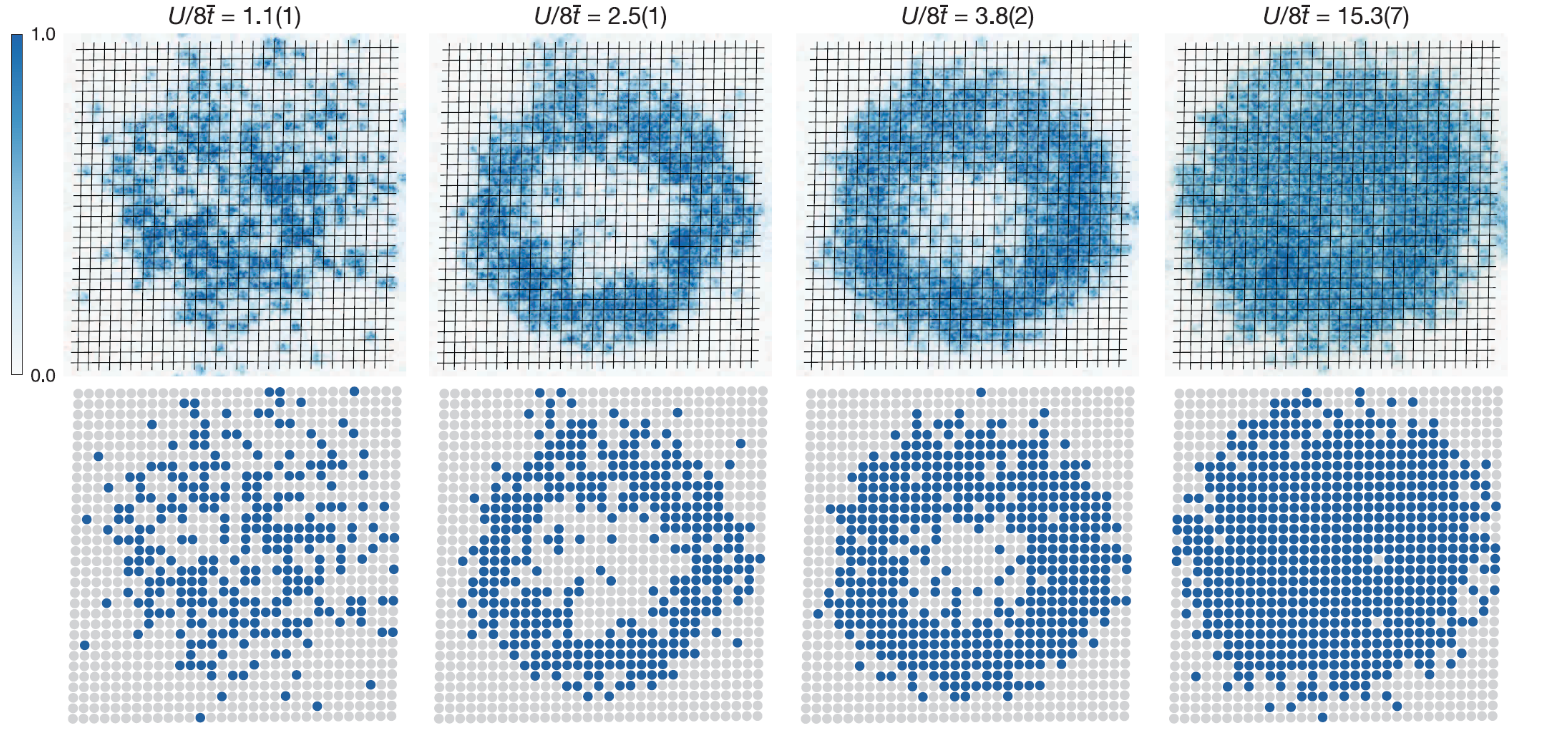}
\caption{\label{fig:microscopeMI}
Observation of the Mott crossover under the quantum gas microscope.
The upper row shows bare high-resolution images of a two-component $^6$Li gas in a two-dimensional optical lattice and a harmonic trap. The lower row show the results of the reconstruction scheme to determine the position of individual atoms within the lattice sites. Blue dots show atoms in singly occupied lattice sites while gray dots correspond either to empty sites or doubly occupied sites (see text).
The interaction strength $U/t$ increases from left to right with values indicated on the figure, while the total atom number increases.
Figure extracted from Ref.~\cite{greif2016}.
}
\end{center}
\end{figure}

Quantum gas microscopes have paved the way to a new generation of studies of the Fermi-Hubbard model.
The Mott insulator has been observed in two-dimensional optical lattice from fluorescence images of $^6$Li~\cite{greif2016} and $^{40}$K~\cite{cheuk2016} atoms. Figure~\ref{fig:microscopeMI} depicts single-site resolved images of the Fermi gas, from which the density distribution, the compressibility, and the density fluctuations, measured at the single-site level, have been extracted. The upper row shows bare images for increasing interaction strength. In this experiment, the spatial resolution is of the order of the lattice spacing. It is sufficient to apply a numerical reconstruction scheme and infer the position of each atom. While the scheme allows for a direct observation of sites with no fermion (no signal, gray dots) or one fermion in the spin $\uparrow$ or $\downarrow$ state (blue dots), strong light-assisted collisions during the imaging process remove all fermions from doubly occupied sites (a notable exception being the Munich fermionic quantum gas microscope~\cite{boll2016}). It is thus not possible to distinguish empty sites from doubly occupied sites. In other words, the imaging method measures the quantity $n_\textrm{det}=\big\langle \hat{n}_{j,\uparrow} + \hat{n}_{j,\downarrow} - 2 \hat{n}_{j,\uparrow}\hat{n}_{j,\downarrow}\big\rangle$.
Remarkably enough, using the relation $\hat{n}_{j,\sigma}^2=\hat{n}_{j,\sigma}$ and the definition of the spin operators~(\ref{spins}), it yields the second moment of the spin operator along the quantization axis,
\begin{equation}
n_\textrm{det}
= \left\langle \left(\hat{n}_{j,\uparrow} - \hat{n}_{j,\downarrow} \right)^2 \right\rangle
= \frac{4}{\hbar^2} \left\langle \left(\hat{S}^z_{j} \right)^2 \right\rangle.
\end{equation}
The images shown on Fig.~\ref{fig:microscopeMI} are then readily interpreted taking this point into account.
For weak interactions (left panel), the gas forms a metallic band conductor and the fermions in the spin $\uparrow$ and $\downarrow$ states spread over almost independently.
When the interaction strength increases (central two panels), a Mott insulator with one atom per site creates in a ring. The exceeding atoms are rejected into the trap center and a peripheral ring. In the center, the low potential energy of the trap favours the accumulation of atoms, which creates a band insulator core with two atoms per site.
The band insulator core is separated from the Mott insulator ring by a narrow metallic ring with significant density fluctuations. In the peripheral ring, the potential energy is larger and only a few sites are populated.
It forms a wedding care structure with coexisting band insulator, metallic, and Mott insulator states.
Finally, for large interactions (right panel), a bulk Mott insulator is formed and the band insulator shrinks.
Adjusting the temperature and the atom number independently by the atomic cloud preparation and the evaporation ramp, it is possible to control the size of the Mott insulator from $N \sim 100$ to $N \sim 700$ atoms~\cite{greif2016,cheuk2016}.

\section{Observation of anti-ferromagnetic order}\label{sec:AForder}

After the demonstration of the fermionic Mott insulator and its characterization in the charge sector, the next challenge has been to turn to the spin sector. Finding evidence of the anti-ferromagnetic phase transition is significantly more difficult than realizing a paramagnetic Mott insulator. On the one hand, it requires to decrease the temperature below the N\'eel temperature, $T_{\textrm{N}} \sim t^2/\kB U$, which, in the strongly correlated regime $\tunnel \ll U$, is typically an order of magnitude smaller than the Mott crossover temperature, $T_{\textrm{M}} \sim U/\kB$. On the other hand, it requires to develop new diagnostic tools, able to probe spin correlation functions, which is not possible with standard absorption imaging techniques. In this section, we review progress in this direction, which provided evidence of spin correlations~(Sec.~\ref{sec:spinorder}) and recently of long-range anti-ferromagnetic order~(Sec.~\ref{sec:AFMorder}).

\subsection{Onset of magnetic spin correlations}\label{sec:spinorder}
The first experiments attempted to reveal precursors of the anti-ferromagnetic phase. Although long-range magnetic order only appears below the critical point, short-range spin correlations are expected at temperatures considerably above the N\'eel temperature. In this section we review the different approaches used to reveal them.

\begin{figure}
\begin{center}
\includegraphics[width=1.0\hsize]{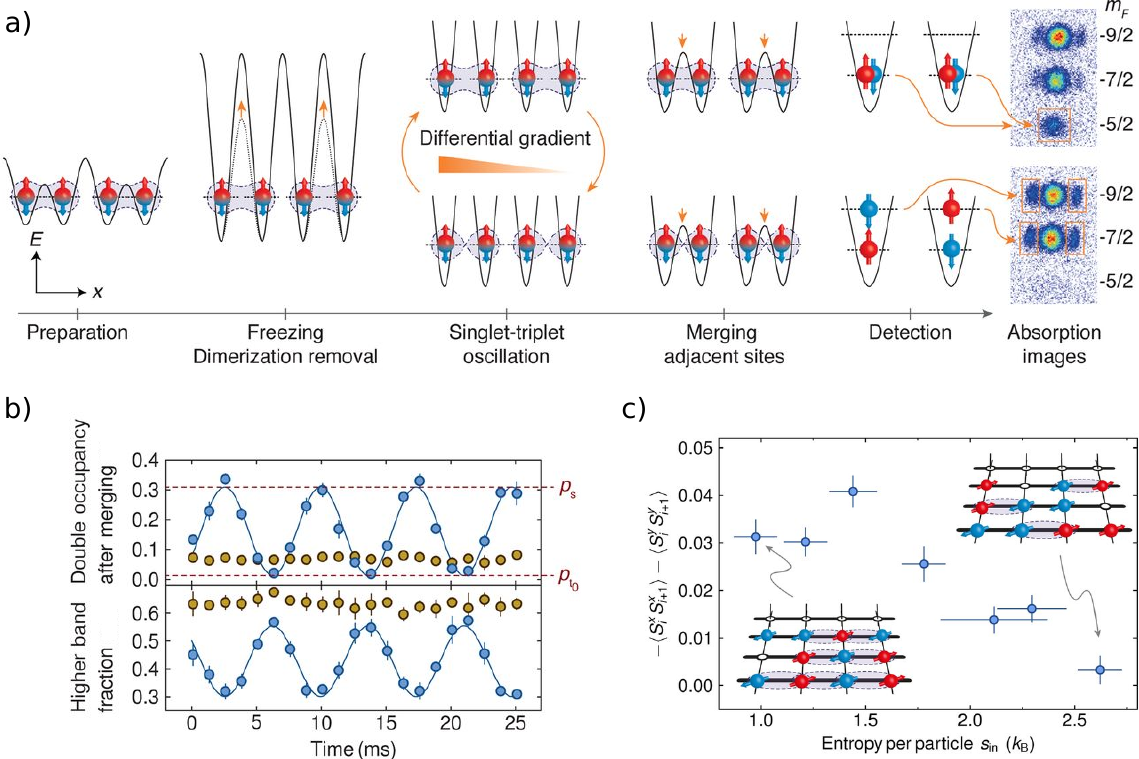}
\caption{\label{fig:spin-corr-merging} Observation of short-range spin correlations in Fermi-Hubbard systems. (a) Sequence exploited to measure nearest-neighbor spin correlation functions by merging adjacent lattice sites using an optical superlattice. After preparing the system of choice (here, a dimerized optical lattice), tunneling is suddenly reduced by raising up the depth of the optical lattice to a very high value. A magnetic field gradient is then exploited to induce oscillations between spin singlet and triplet states of two adjacent lattice sites. Subsequently, the lattice structure is modified in order to merge them into a single site. Singlet states lead to the formation of a doubly occupied site in the motional ground state of the well, whereas triplet states lead to the promotion of one atom to a higher band. They can be detected using a rf transfer to the ancilla state $m_F=5/2$ and band mapping, respectively. (b) Singlet-triplet oscillations in a dimerized lattice. Blue (ocre) points correspond to the merging of sites connected by strong (weak) tunneling. In the first case, the presence of an imbalance between the singlet and triplet populations leads to an oscillation of finite amplitude, which can be used to infer the magnetic correlations of the system. (c) Measurement of the nearest-neighbor spin correlation function in an anisotropic simple cubic lattice as a function of the entropy per particle (related to the temperature) of the system measured by site merging. The various panels are adapted from Ref.~\cite{greif2013}.
}
\end{center}
\end{figure}

\paragraph*{Merging of adjacent lattice sites~--}
The first experiment to reveal the emergence of short-range spin correlations was performed in Zurich~\cite{greif2013}. After preparation of a Fermi-Hubbard system using the $|9/2,-9/2\rangle$ and $|9/2,-7/2\rangle$ spin states of $^{40}$K at various values of the ratio $U/t$, the dynamics of the system was frozen by raising up a very deep optical lattice. Neighboring lattice sites were then merged two-by-two exploiting an optical superlattice. In this scheme, the orbital state of the atoms after the merging allows to identify whether the two atoms were initially in a spin singlet state $|s\rangle=\left(|\uparrow,\downarrow\rangle-|\downarrow,\uparrow\rangle\right)/\sqrt{2}$ or in the triplet state $|t_0\rangle=\left(|\uparrow,\downarrow\rangle+|\downarrow,\uparrow\rangle\right)/\sqrt{2}$, where $\vert\sigma_1,\sigma_2\rangle$ denotes the spin state in the left and right of the pair of sites. Because the two-particle fermionic wavefunction needs to be antisymmetric, in the first case the merging will yield a doubly occupied site, whereas in the second case one of the fermions will be promoted to a higher band. These situations can then be distinguished experimentally using rf-spectroscopy of doubly occupied sites and band-mapping, respectively, see Fig.~\ref{fig:spin-corr-merging}. When performing global measurements over the system it is advantageous to induce in addition oscillations between the singlet and the triplet state before the merging. The oscillations can be obtained by creating an energy bias between the left and right wells using a magnetic field gradient. The amplitude of the resulting singlet-triplet oscillations then directly yields the imbalance between the singlet and triplet populations of the complete system. Interestingly, these correspond to the nearest-neighbor spin-spin correlation function
\begin{equation}
\langle \hat{S}^x_j \hat{S}^x_{j+1}\rangle+\langle \hat{S}^y_i \hat{S}^y_{j+1}\rangle=-\left(p_s-p_{t_{0}}\right)/2,
\end{equation}
where $p_s$ and $p_{t_0}$ denote the fraction of atoms forming singlets and triplets $t_0$ respectively.
In a first series of experiments, this method was used to show the existence of sizeable short-range correlations in dimerized and in anisotropic simple cubic lattices, where the tunnelling was much stronger along one lattice direction. Comparisons with DMRG simulations~\cite{sciolla2013} and with a cluster version of DMFT~\cite{imriska2014} allowed to determine the temperature of the system. Subsequent studies were performed in different dimensions and lattice geometries, showing in all cases short-range spin correlations~\cite{greif2015}. Exploiting a periodic modulation of the lattice position in time allowed as well to control independently the strength of the tunnelling and exchange energies, and to enhance or even change the sign of the magnetic correlations in the system, from anti-ferromagnetic to ferromagnetic~\cite{goerg2018}. Finally, measurements of nearest-neighbor spin correlations have been recently performed as well in $SU(6)$ systems, yielding analogous results~\cite{ozawa2018}.

\begin{figure}
\begin{center}
\includegraphics[width=1.0\hsize]{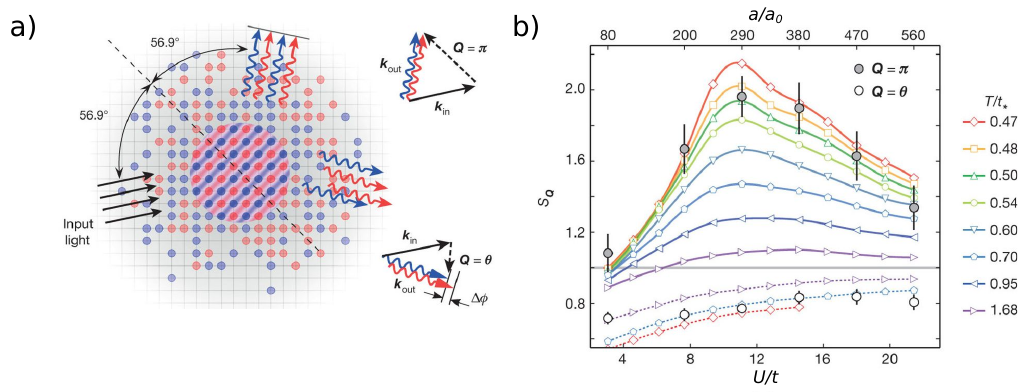}
\caption{\label{fig:spin-corr-Bragg} Observation of spin correlations in the Fermi-Hubbard model using Bragg scattering of light. (a)~Sketch of the experimental scheme. Near-resonant light (black straight arrows) is sent onto the atomic sample. The light that is coherently scattered by the two spin states (red and blue wiggly arrows) is collected along two directions. When the momentum transferred to a scattered photon is $\mathbf{Q}=\boldsymbol{\pi}=(-\pi/a,-\pi/a,\pi/a)$, the Bragg condition for scattering from an antiferromagnet is fulfilled and coherent interference occurs. This does not happen for most angles $\mathbf{Q}=\boldsymbol{\theta}$. (b)~Magnetic structure factor $S_{\mathbf{Q}}=S_s^z(\mathbf{Q})$ measured experimentally. When reducing the temperature of the system $T/t$, the short-range magnetic correlations that precede antiferromagnetic order become visible for $\mathbf{Q}=\boldsymbol{\pi}$. They are maximum for intermediate values of $U/t$, as expected from the phase diagram. Solid lines correspond to determinant quantum Monte-Carlo simulations of the system, which are in good agreement with the experimental results. Figure adapted from Ref.~\cite{hart2015}.
}
\end{center}
\end{figure}

\paragraph*{Bragg scattering of light~--} An alternative method to probe spin correlations in the three-dimensional Fermi-Hubbard model, analogous to X-ray or neutron scattering in solid-state systems, was subsequently demonstrated by the Rice group~\cite{hart2015}. In these experiments, realized with a gas of $^6$Li atoms in the spin states $|1/2,1/2\rangle$ and $|1/2,-1/2\rangle$, the lattice system was illuminated with a near-resonant laser beam, and the coherent scattering of light from the sample was recorded with a CCD camera. In three-dimensions, the incoming and reflected beams interfere constructively for certain angles and scattering is dramatically enhanced compared to the other angles~\cite{miyake2011}, see Fig.~\ref{fig:spin-corr-Bragg}(a)\footnote{Note that in one and two dimensions diffractive light scattering occurs at any angle of incidence, as experimentally demonstrated in  Ref.~\cite{weitenberg2011}.}. Because antiferromagnetic order effectively doubles the lattice spacing of each spin component, the presence of enhanced scattering at certain angles when using a beam that interacts differently with the two spin states reveals the presence of magnetic correlations in the system~\cite{corcovilos2010}. Experimentally, this requirement is fulfilled by setting the beam detuning in between the optical transition frequencies of the two states. The intensity of the scattered light is then recorded for two angles, one satisfying the constructive interference condition and one away from it. The ratio of these two measurements allows to determine the so-called spin structure factor of the system, which characterizes the Fourier spectrum of the spatial magnetic correlations, averaged over the complete system. It is given by
\begin{equation}\label{SpinStructFact}
S^z_{\textrm{s}} \big({\vec{q}}\big) = \frac{1}{N}\sum_{\vec{r}} C_s^z(\mathbf{r}) \exp \left(-i \vec{q} \cdot \vec{r} \right),
\end{equation}
where $\mathbf{q}$ corresponds to the momentum difference between the incoming and outgoing beams (which can be selected by the observation angle), $N$ is the number of lattice sites, and the spatial spin correlation function $C_s^z(\mathbf{r})$ is defined in Eq.~(\ref{eq:SpinCorrFunct}). The magnetic structure factor is maximized at the corners of the first Brillouin zone, $\mathbf{q}=(\pm \pi/a,\pm \pi/a,\pm \pi/a)$, which was thus the value probed in the experiment. The results obtained as a function of $U/t$ and temperature are displayed on Fig.~\ref{fig:spin-corr-Bragg}(b), and show that for sufficiently low temperatures magnetic correlations appear in the system. The magnetic correlations are maximized for an intermediate value of $U/t\sim12$, reflecting the fact that the N\'eel temperature is maximal for intermediate coupling strengths (see Sec.~\ref{sec:FHmPhasDiag}).

\paragraph*{Correlation analysis of in situ images~--}
In two-dimensional systems, the spatial spin correlations and the magnetic structure factor can be extracted locally from the \emph{in situ} images. As explained in Sec.~\ref{sec:2D}, most quantum gas microscopes measure the density distribution of the system without being able to distinguish between the $\uparrow$ and $\downarrow$ spin states. It is, however, possible to obtain spin information by selectively eliminating one or the other spin component, and detecting the remaining spin distribution of the system. The experiment is performed in the following way: After freezing the atomic distribution by suddenly increasing the lattice depth, the atoms in one of the spin states are ejected using a resonant laser pulse without affecting the atomic distribution of the other spin component. This is possible due to the large Zeeman shift existing between the two optical transitions at the magnetic field where the experiments are performed. In-situ imaging then maps the detected on-site population $\hat{n}_j^{\textrm{det}}$ onto the local spin $\hat{S}^z_j = \hbar \big(\hat{n}_j^{\textrm{det}}-1/2\big)$. By combining density-density correlations measured in images where only $\uparrow$ atoms or $\downarrow$ atoms were removed, and an image where no removal was performed, one can compute the local spin
correlation function $C_s^z(\mathbf{r})$ defined in Eq. (\ref{eq:SpinCorrFunct}). The Harvard and MIT groups performed such experiments using spin-balanced systems of $^{6}$Li and $^{40}$K atoms respectively~\cite{parsons2016,cheuk2016b}, whereas the Princeton group focused instead on the situation where the spin populations were imbalanced~\cite{brown2017}. The latter is equivalent to the introduction of an effective Zeeman field $h=\left(\mu_{\uparrow}-\mu_{\downarrow}\right)/\hbar$, where $\mu_{\uparrow,\downarrow}$  denote the chemical potentials of the two spin components. This term modifies the Fermi-Hubbard Hamiltonian, Eq.~(\ref{FH-Hamiltonian}), by adding a term $-h\sum_j \hat{S}^z_j$. It also breaks the $SU(2)$ symmetry of the model, and leads to canted anti-ferromagnetic order. Indeed, this allows to accommodate magnetization along the field direction, whereas the system can still reduce its energy via super-exchange processes by ordering anti-ferromagnetically in the perpendicular plane.

The imaging scheme used in the Munich quantum gas microscope is slightly different because, in order to fix the position of the atoms during the imaging process, it exploits an imaging pinning lattice of spacing smaller than the "physics" lattice where the experiments are performed. This allows to avoid light-assisted collisions during the imaging process and detect the doubly occupied lattice sites~\cite{omran2015}. The oversampling of the "physics" lattice by the pinning lattice also allows to spatially separate the two spin components in one-dimensional systems before imaging them. To this end, a magnetic field gradient is applied along a direction orthogonal to the lattice. This scheme thus gives access to single-shot spin-resolved images, from which the calculation of spin correlation functions becomes straightforward~\cite{boll2016}. This method has been exploited to study phenomena such as spin-charge separation~\cite{hilker2017} and incommensurable magnetism in Hubbard chains~\cite{salomon2018}.

In experiments with more moderate optical resolution, it is still possible to extract the spatial spin-correlation functions from a correlation analysis of the \emph{in situ} density distributions of the $\uparrow$ and $\downarrow$ states, as demonstrated by the Bonn group~\cite{drewes2016b}. These measurements yield the spin structure factor of the system at zero momentum $S^z_s(\mathbf{q=0})$. However, the spin correlations can be coherently manipulated by letting the system evolve in the presence of a magnetic field gradient in a Ramsey-type sequence. Using this method, periodic spin patterns with well defined momentum can be imprinted onto the gas, and exploited to probe the spin structure factor at various momenta~\cite{wurz2017}.

\subsection{Observation of the long-range anti-ferromagnet}\label{sec:AFMorder}

All the experiments discussed in the previous section observed the short-range spin correlations that precede the anti-ferromagnetic phase. However, the temperature of the system was still above the critical N\'eel temperature below which anti-ferromagnetic long-range order appears.

\paragraph*{Cooling down below the N\'eel temperature~--}
Lowering further the temperature, down below the N\'eel scale $T_\textrm{N}$, in an atomic Fermi gases is a major experimental challenge.
The gas is initially cooled in a harmonic trap \emph{via} forced evaporation, resulting in the typical level of quantum degeneracy $T/\TF\sim0.05$, where $\TF$ is the Fermi temperature. This has been sufficient to reach the Mott insulator regime as discussed above.
When entering the quantum degenerate regime, the thermalization slows down owing to the Pauli blocking of collisions and to inelastic processes, which lead to hole heating~\cite{carr2004} and make cooling beyond this value hardly possible.
The gas is subsequently loaded into the optical lattice potential adiabatically, hence conserving the total entropy.
Due to the inhomogeneous trapping, the entropy is, however, inhomogeneously distributed within the system.
The metallic shells store a much larger amount of entropy than the Mott insulating cores.
This typically leads to temperatures in the half-filled Mott insulator region of the order of $T \sim 0.5 \tunnel/\kB$,
which is only approximately $40\%$ above the N\'eel temperature for the typical value $\tunnel \simeq U/8$~\cite{hart2015}.
Such a low temperature was reached in a compensated lattice, which creates a quartic, rather than harmonic, trap~\cite{mathy2012}.
One can further take advantage of the inhomogeneous density distribution within the trap and further cool the system. The simplest approach, suggested in Ref.~\cite{ho2009b}, is to shape adequately the trapping potential to enlarge the size of the metallic regions where the entropy per site is maximal. This region hence acts as a reservoir of entropy and permits to reduce the global temperature of the system. At the same time, a central half-filled region is required in order to observe anti-ferromagnetic order. This entropy redistribution scheme, combined with a fermionic quantum-gas microscope, has been implemented successfully in 2017 in Ref.~\cite{mazurenko2017}. By decreasing the temperature of the gas approximately a factor of two, it allowed to finally observe the transition to a long-range ordered Fermi-Hubbard anti-ferromagnet, as we discuss now.

\paragraph*{Observation of long-range anti-ferromagnetic order~--}
The experiment of Ref.~\cite{mazurenko2017} has been implemented on the two-dimensional Hubbard model. The latter allows full exploitation of the quantum-gas microscope~\cite{mazurenko2017}. However, in two dimensions, the thermal fluctuations suppress the long-range correlations and strict long-range magnetic order can show up only at zero temperature, see discussion in Sec.~\ref{sec:SpinSector}. For any finite temperature,  the spin correlation function~(\ref{eq:SpinCorrFunct}), decays exponentially,
\begin{equation}\label{eq:StaggeredSpinCorr}
C_{\textrm{s}}(\vec{r}) \sim (-1)^{\vec{r}} \exp \left(- \vert\vec{r}\vert / \xi_{\textrm{s}}\right),
\end{equation}
where $\xi_{\textrm{s}}(T)$ is the staggered-spin correlation function,
hence producing, in principle, only short-range magnetic order. However, in a finite system such as ultracold atoms, an effective long-range order is found as long as the correlation length exceeds the system size.
The observation of such an effective long-range order has been a \textit{tour de force}.
In the experiment of Ref.~\cite{mazurenko2017}, a flat two-dimensional potential is created using a digital micromirror device to compensate the harmonic trapping induced by the lattice and create sharp edges. This creates a nearly perfect flat disk of about 10 sites in diameter in the center of the sample, totalizing about $75$ lattice sites. Lattice site-resolved imaging with results such as those shown on Fig.~\ref{fig:microscopeMI} allows to check that a Mott insulator state with almost exactly one atom per site is realized.

\begin{figure}
\begin{center}
\includegraphics[width=1.0\hsize]{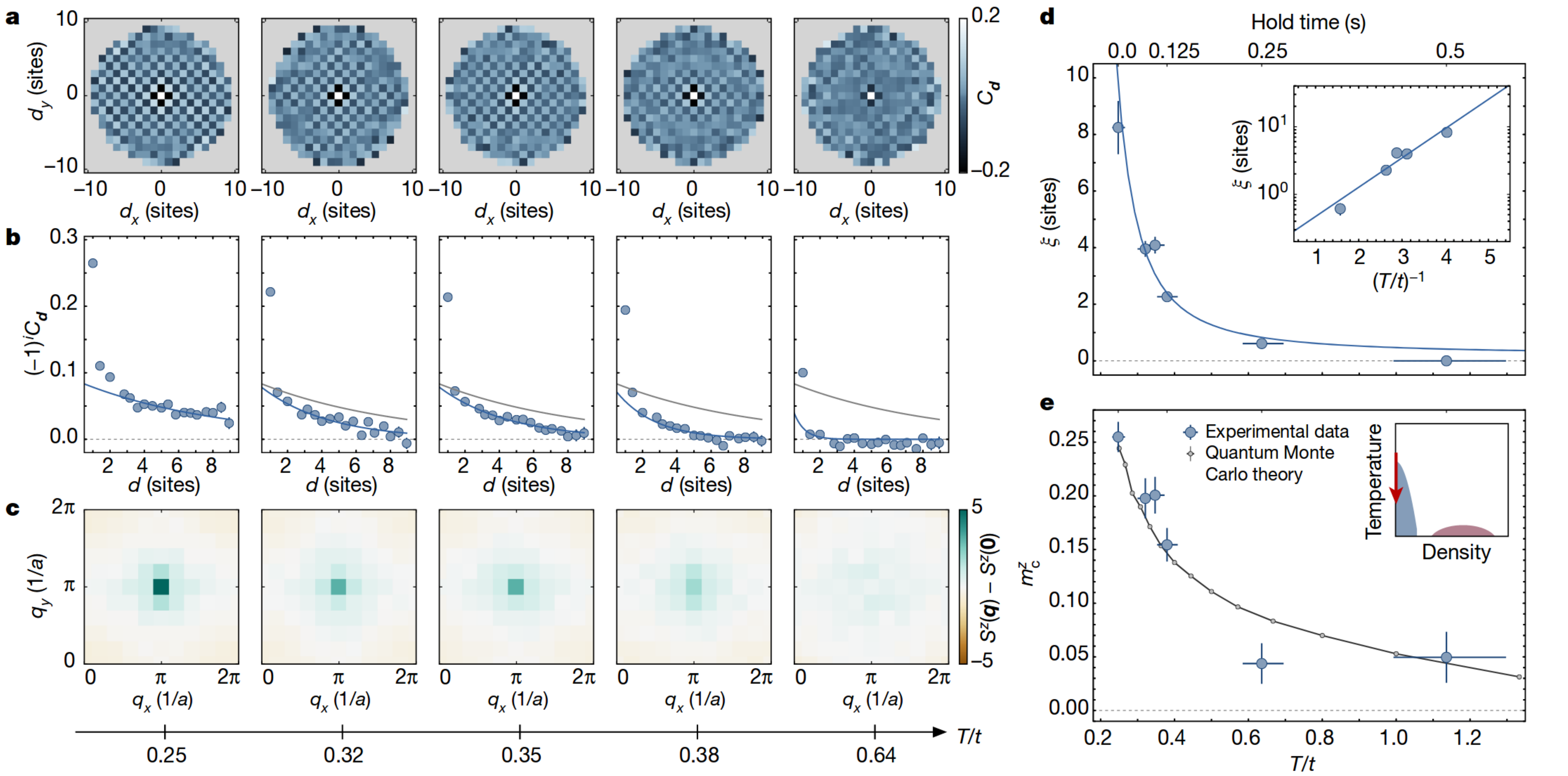}
\caption{\label{fig:AFM}
Observation of the anti-ferromagnetic phase in the two-dimensional Hubbard model.
(a)~Two-dimensional map of the spin correlation function in a 10-wide Mott insulator disk as computed from site-resolved, spin-selective images.
(b)~Staggered spin correlation function versus the radial distance. The solid blue line is an exponential fit to the experimental data. The solid gray lines on the four right panels reproduce the fitted curve obtained at the lowest temperature (left panel, $T \simeq 0.25 t/\kB$) for reference.
(c)~Spin structure factor in quasi-momentum space.
The various panels on Figs.~(a), (b), and (c) correspond to an increasing temperature, indicated on the lower axis.
(d)~Staggered spin correlation length as a function of the temperature, as found from the fits shown on panel~(b). The solid line is an fit of Eq.~(\ref{eq:SpinCorrLength}) to the data.The inset shows the same quantity in semi-log scale versus the inverse of the temperature.
(e)~Staggered magnetization versus the temperature as found in the experiment (blue points) and quantum Monte Carlo calculations (gray points and line).
Figure extracted from Ref.~\cite{mazurenko2017}.
}
\end{center}
\end{figure}

Figures~\ref{fig:AFM}(a), (b), and (c) show the experimental results for, respectively,
the two-dimensional spin correlation function $C_{\textrm{s}}(\vec{r})$~[Eq.~(\ref{eq:SpinCorrFunct})],
the radial staggered-spin correlation function $(-1)^{\vec{r}}C_{\textrm{s}}(r)$~[Eq.~(\ref{eq:StaggMagn})] as a function of the radius,
and the spin structure factor $S_{\textrm{s}}(\vec{r})$~[Eq.~(\ref{SpinStructFact})].
The various panels correspond to an increasing temperature, from left to right.
The two-dimensional correlation function, Fig.~\ref{fig:AFM}(a), shows a clear checkerboard structure, alternating between positive (dark blue) and negative (white) values.
While blurred at the largest temperature ($T \simeq 0.64\tunnel/\kB$), this structure raises up when the temperature lowers down to the smallest attained value ($T \simeq 0.25\tunnel/\kB$).
This is consistent with the appearance of a staggered magnetization and finite-range anti-ferromagnetism.
The onset of an effective long-range anti-ferromagnetic order is further characterized by the staggered correlation function, see Fig.~\ref{fig:AFM}(b).
The decay of $(-1)^{\vec{r}}C_{\textrm{s}}(\vec{r})$ as a function of the radial distance $\vert\vec{r}\vert$ between the spins is well fitted to an exponential function, beyond the nearest-neighbor site. This is consistent with the expected finite-temperature behavior, Eq.~(\ref{eq:StaggeredSpinCorr}), and permits the direct measurement of the staggered spin correlation length $\xi_{\textrm{s}}$. The results of the fits are plotted on Fig.~\ref{fig:AFM}(d) as a function of the temperature. The correlation length is smaller than one lattice site for large temperature and sharply increases when the temperature decreases. For the lowest temperature, it reaches $\xi_{\textrm{s}} \simeq 8$ in units of the lattice spacing, that is about the total size of the Mott insulator core. This confirms the onset of the effective long-range anti-ferromagnetic order.
Moreover, the spin correlation length scales exponentially with the temperature, see Inset of Fig.~\ref{fig:AFM}(d). It is consistent with the theoretical prediction
\begin{equation}\label{eq:SpinCorrLength}
\xi_{\textrm{s}}\left(T\right) \propto \exp \left( 2\pi \rho_{\textrm{s}} / \kB T \right),
\end{equation}
where $\rho_{\textrm{s}}$ is the spin stiffness~\cite{manousakis1991}. The value of the latter, extracted from fits to the experimental data, yields the value $\rho_{\textrm{s}} \simeq 1.6 \tunnel$, which agrees with the theoretical prediction of the spin-wave theory, $\rho_{\textrm{s}} \simeq 1.3 \tunnel$~\cite{denteneer1993}, within about $20\%$.

Other quantities extracted from the same site-resolved, spin-selective imaging provide further signatures of the anti-ferromagnetic order.
For instance, the spin structure factor, displayed on Fig.~\ref{fig:AFM}(c), shows a marked peak at the corners of the first Brillouin zone, $\vec{q} = \big(\pm\pi/a,\pm\pi\big/a)$. It is consistent with the onset of the checkerboard structure observed on Fig.~\ref{fig:AFM}(a) when the temperature decreases. The staggered magnetization along the quantization axis $z$ is displayed on Fig.~\ref{fig:AFM}(e). It increases from $\mathcal{M}^z \simeq 0.05$ for the largest temperature up to $\mathcal{M}^z \simeq 0.25$ for the smallest one, in excellent agreement with quantum Monte Carlo calculations performed on a $10\times 10$ square lattice.
It is about $50\%$ off the zero-temperature prediction, $\mathcal{M}^z \simeq 0.35$ in two-dimensions.
Moreover, the images reveal the $SU(2)$ symmetry of the anti-ferromagnetic phase. Since the staggered spin-ordering vector points in random directions, shot-to-shot images display large fluctuations, again in agreement with quantum Monte Carlo calculations.

\section{Outlook and other research directions}\label{sec:outlook}
The dramatic progress realized in the last few years on the quantum simulation of the Fermi-Hubbard model opens a plethora of exciting new perspectives.
An immediate one is to explore the full phase diagram depicted on Fig.~\ref{fig:FHmPhaseDiag}.
So far, the anti-ferromagnetic phase has only been observed in the strongly interacting (Mott) regime, $U \gg \tunnel$. The weakly interacting (Slater) regime, $\tunnel \ll U$, which may be typical of some solid-state materials described by the Fermi-Hubbard model, is even more challenging. There, the critical temperature scales exponentially with the inverse interaction strength $\tunnel/U$ [see Eq.~(\ref{eq:NeelTSlater})] and thus requires the development of new cooling schemes in ultracold-atom systems.
Another major challenge is to extend the phase diagram away from half filling.
In the context of strongly-correlated electronic systems, this is realized by doping the material thanks to the substitution of a fraction of some constitutive atomic species by another species. This may bring an excess of electrons (electron doping) or a defect of electrons (hole doping), depending on the substitution.
Compared to the half-filling case, the theoretical investigation of the exact doped Fermi-Hubbard model is much more demanding. This owns to the increasing difficulty to numerically simulate the system for relevant parameters. For instance, quantum Monte Carlo simulations of the doped Fermi-Hubbard model are limited by the exponential scaling of the sign problem with doping, inverse temperature, and interaction strength~\cite{staar2013}.
The dynamical mean field theory (DMFT) is among the privileged approaches to strongly-correlated materials. It was proved to be exact in the limit of infinite connectivity, i.e.\ in infinite dimension, and to successfully simulate a wide variety of three-dimensional materials~\cite{georges1992,georges1996}. However, its validity in low-dimensional materials is limited.

\begin{figure}
\begin{center}
\includegraphics[width=0.95\hsize]{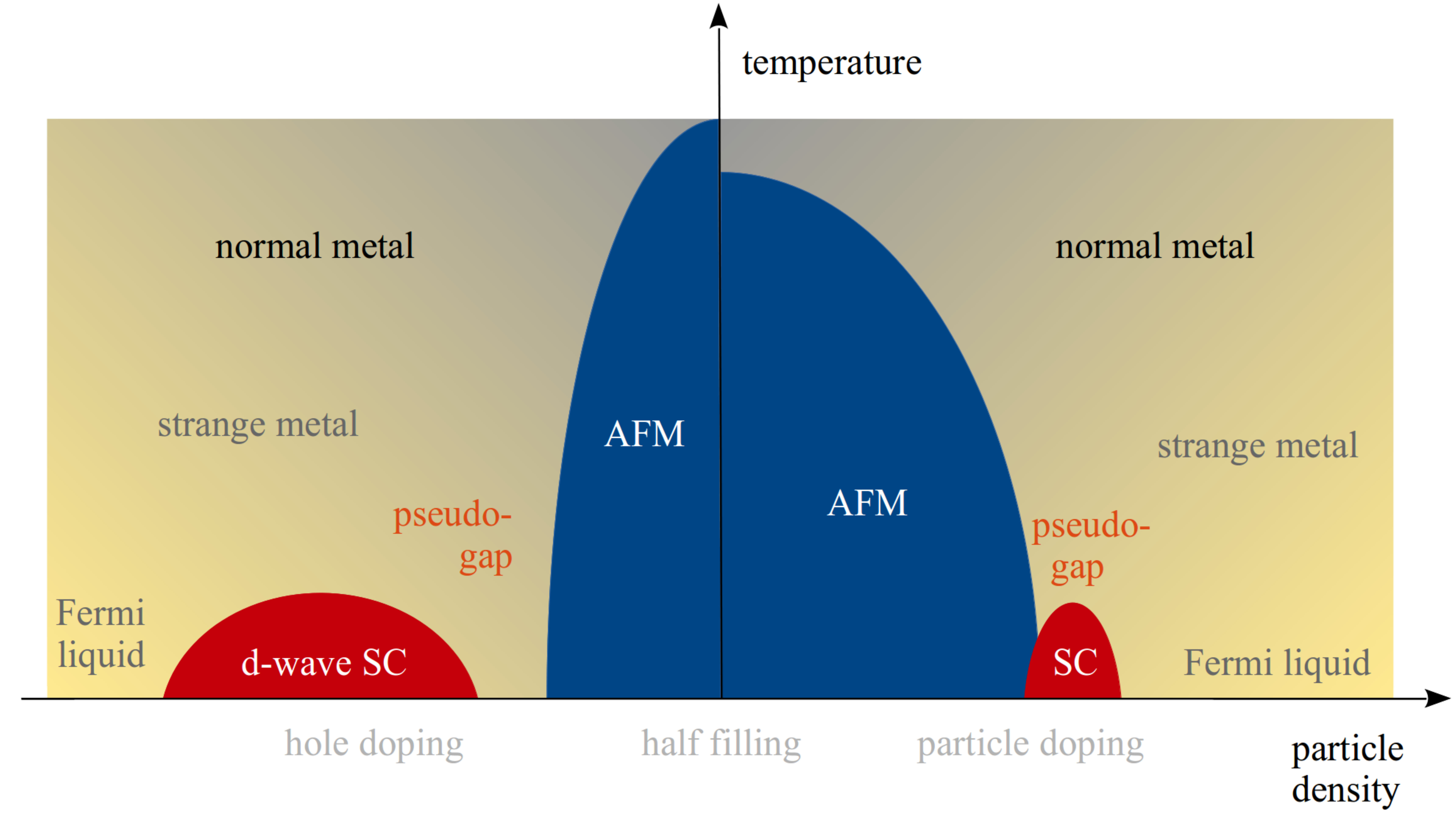}
\caption{\label{fig:DopedPhaseDiagram}
Schematic phase diagram of the doped Fermi-Hubbard model.
The central vertical axis represents the half-filled case discussed in the core of the present review.
The horizontal axis represents the particle density and comprises the particle doping sector, corresponding to an excess of particles (right-hand side), and the hole doping sector, corresponding to the defect of particles (left-hand side).
In each sector, the diagram contains three phases, namely
(i)~the anti-ferromagnetic (AFM) Mott insulator phase (blue),
(ii)~the $d$-wave superconducting (SC) phase (red),
and (iii)~the normal metallic phase (yellow-grey).
In the hole-doping sector, the normal metal comprises several regions including the normal Fermi liquid, the strange metal, and the pseudo-gap regimes. The latter are separated by smooth crossovers.
}
\end{center}
\end{figure}

The observation of high-Tc superconductivity in cuprates in the mid-1980's~\cite{bednorz1986} has triggered enormous interest for the physics of doped Mott insulators. The high value of the critical temperature and the fact that it was observed in compounds where superconductivity was not expected suggested the existence of a new mechanism for superconductivity.
The extensive experimental investigation of cuprates permitted to establish the universal phase diagram depicted on Fig.~\ref{fig:DopedPhaseDiagram}. High-Tc superconductivity is just one aspect of the very unusual features of cuprates. The first striking feature of the phase diagram is the breaking of the particle-hole symmetry. The particle-doping sector (right-hand side of the diagram), and the hole-doping sector (left-hand side of the diagram), both comprise an anti-ferromagnetic Mott insulating phase (AFM), a superconducting phase (SC), and a normal metal phase. However, the anti-ferromagnetic phase is more extended in the particle-doping sector than in the hole-doping sector. Conversely, the superfluid phase is more extended in the hole-doping sector than in the particle-doping sector. The order parameter of the superconducting phase has $d$-wave symmetry.
The so-called "normal metal" phase is actually not so "normal" and presents several unusual properties. For instance, the region right above the superconductor dome and for moderate doping exhibits abnormally few states and is termed the \textit{pseudo-gap} regime. For higher temperature, the pseudo-gap vanishes but the metal resistivity shows an usual linear temperature dependence in the so-called \textit{strange metal} regime. Finally, for stronger doping, a more usual quadratic temperature dependence, characteristic of a Fermi liquid is recovered.
The boundaries between these regimes of the normal metal phase are smooth and should be regarded as crossovers.

While this universal phase diagram is consensual, the identification of the minimal model able to display such a rich diagram, hosting in particular high-Tc superconductivity and strange metallic behavior, remains debated. The two-dimensional Fermi-Hubbard model appears to be a promising candidate, although the mechanism leading to a superconducting phase is still the object of active research~\cite{lee2006,proust2018}.
In view of the recent developments, quantum simulators based on ultracold atoms appears particularly suited to progress on this question.
Doping the system by controlling the density of the bulk, a recent experiment has observed the progressive suppression and eventually the cancellation of the long-range anti-ferromagnetic order~\cite{mazurenko2017}.
This is compatible with the emergence of the pseudo-gap phase, which is expected to be the precursor of the superconducting phase.
Although the present experiments are performed at temperatures about one order of magnitude larger than the critical temperature for superfluidity, they already give access to the high-temperature sector of this phase diagram where anomalous transport properties and hidden anti-ferromagnetic order, characterized by non-local string correlation functions, are expected. Generalizing the measurements performed in one-dimensional systems~\cite{hilker2017} to the two-dimensional case, these could be detected in the current experiments. Decreasing the temperature and increasing the doping, the formation of the d-wave superconducting phase is now within reach.
The temperature of the system could be decreased by implementing more elaborate entropy redistribution schemes, see for instance Refs.~\cite{bernier2009,ho2009b,grenier2014}. For instance, by creating a band-insulating phase, where the entropy per site vanishes, isolating it from the higher-entropy regions of the cloud and finally doubling the lattice spacing, the ultra-low temperatures required to observe superfluidity might be accessible.
The first step of this protocol has recently been implemented experimentally~\cite{chiu2018}.

Another promising research direction, accessible at the currently available temperatures, is the investigation of transport properties. Whereas early experiments probing the bulk transport properties of the system were typically performed far from the linear response regime~\cite{ott2004,strohmaier2007,schneider2012}, more recent ones~\cite{anderson2017} reach this limit and relate the measured transport observables to the thermodynamic properties of the system.
Furthermore, diffusion measurements of the repulsive Fermi-Hubbard model have been recently exploited to characterize density and spin transport~\cite{brown2018,nichols2018}. Interestingly, at low temperatures and away from half-filling, the experiments reveal anomalous particle transport and a linear dependence of the resistivity of the system with temperature, analogous to the strange metal behaviour observed in the cuprates~\cite{xu2016,brown2018}.

Finally, other promising directions for future research include the study of out-of-equilibrium dynamics in this system~\cite{strohmaier2010,schneider2012,pertot2014,scherg2018}, and the interplay of strong correlations with disorder~\cite{lsp2010,kondov2015,schreiber2015} and topology \cite{jotzu2014,mancini2015,flaschner2016,nakajima2016}.

\bibliographystyle{naturemag}


\end{document}